\documentclass[preprint2]{proto}
\usepackage{times}

\newcommand{\refs}{\par\noindent\hangindent=1pc\hangafter=1}
\def\teff{T_{\rm eff}}
\def\gcc{\rm g.cm^{-3}}

\def\simgr{\,\hbox{\hbox{$ > $}\kern -0.8em \lower 1.0ex\hbox{$\sim$}}\,}
\def\simle{\,\hbox{\hbox{$ < $}\kern -0.8em \lower 1.0ex\hbox{$\sim$}}\,}
\def\msolyr{M_\odot/{\rm yr}}

\def\msol{M_\odot}
\def\mjup{M_{\rm Jup}}
\def\mnep{M_{\rm Nep}}
\def\rjup{R_{\rm J}}
\def\mearth{\,{\rm M}_\oplus}
\def\mnep{\,{\rm M}_{\rm Nep}}

\def\menv{M_{\rm env}}

\def\mz{M_{\rm Z,env}}

\def\te{T_{\rm eff}}

\def\simgr{\,\hbox{\hbox{$ > $}\kern -0.8em \lower 1.0ex\hbox{$\sim$}}\,}
\def\simle{\,\hbox{\hbox{$ < $}\kern -0.8em \lower 1.0ex\hbox{$\sim$}}\,}
\def\beq{\begin{equation}}
\def\eeq{\end{equation}}

\begin{document}

\title{\textbf{\LARGE Gaseous Planets, Protostars And Young Brown Dwarfs :\\ Birth And Fate}}

\author {\textbf{\large G. Chabrier, I. Baraffe, F. Selsis}} 

\affil{\small\em Ecole Normale Sup\'erieure de Lyon} 

\author {\textbf{\large T. S. Barman}} 

\affil{\small\em Department of Physics and Astronomy, UCLA} 

\author {\textbf{\large P. Hennebelle}} 

\affil{\small\em Ecole Normale Sup\'erieure, Paris}

\author {\textbf{\large Y. Alibert}} 

\affil{\small\em Physikalisches Institut, University of Bern} 

\begin{abstract}
\baselineskip = 11pt
\leftskip = 0.65in   
\rightskip = 0.65in  
\parindent=1pc
{\small We review recent theoretical progress aimed at understanding the formation and the early stages of evolution of giant planets, low-mass stars and brown dwarfs. Calculations coupling giant planet formation,
within a modern version of the core accretion model that includes planet migration and disk evolution, and subsequent evolution yield consistent determinations of the planet structure and evolution.
Uncertainties in the initial conditions, however, translate into large uncertainties in the luminosity at early stages. It is thus not possible to say whether young planets are faint or bright compared with low-mass young brown dwarfs. We review the effects of irradiation and evaporation on the evolution of short period planets and argue that substantial mass loss may have occurred for these objects.

Concerning star formation, geometrical effects in protostar core collapse are examined by comparing 1D and 3D calculations. Spherical collapse is shown to significantly overestimate the core inner density and temperature and thus to yield incorrect initial conditions for pre-main sequence or young brown dwarf evolution. Accretion is also shown to occur non-spherically over a very limited fraction of the protostar surface. Accretion affects the evolution of young brown dwarfs and yields more compact structures for a given mass and age, thus
fainter luminosities, confirming previous studies for pre-main sequence stars. This can lead to severe
misinterpretations of the mass and/or age of
young accreting objects from their location in the HR diagram. Since accretion
covers only a limited fraction of the protostar surface, we argue that newborn stars and brown dwarfs should appear rapidly over an  extended area in the HR diagram, depending on their accretion history, rather than on a well defined birth line. Finally, we suggest that
the distinction between planets and brown dwarfs be based on an observational diagnostic, reflecting
the different formation mechanisms between these two distinct populations, rather than on an arbitrary, confusing definition.   \\~\\~\\~}

\end{abstract}  

\section{\textbf{INTRODUCTION}}

One of the fundamental questions of astrophysics remains the characterization of the formation of planets and stars. The mass ranges of the most massive planets and of the least massive brown dwarfs certainly
overlap in the $\sim 1$-10 $\mjup$ range; it is thus interesting to explore our understanding of the
planet and star formation mechanisms in a common review.

The growing number of discovered extrasolar giant planets, ranging now from neptune-mass
to few jupiter-mass objects, has questioned our understanding of planet formation and evolution. The significant fraction of exoplanets in close orbit to their parent star, in particular, implies a revision of our
standard scenario of planet formation. Indeed, these objects are located well within the so-called ice line and could not have formed in-situ.
This strongly favors planet migration
as a common process in planet formation. This issue is explored in \S \ref{planet} where we present consistent calculations between a revised version of the core accretion
model, which does take planet migration into account, and subsequent evolution. In this section, we also review our
current understanding of the effects of irradiation and evaporation on the evolution of short-period planets, hot-neptunes and hot-jupiters, and review present uncertainties in the determination
of the evaporation rates. In \S \ref{collapse}, we briefly review our current understanding of protostellar core collapse and we show that non-spherical
calculations are required to get proper accretion histories, densities and thermal profiles for the prestellar core. The effect of accretion on the early contracting phase of pre-main sequence stars and young brown dwarfs, and a review of observational determinations of accretion rates, are considered in \S \ref{accretion}.
Finally, through out this review, we have adopted as the definition of {\it planet} an object formed by the three-step process
described in \S \ref{planet_formation}, characterized by a central rocky/icy core built by accretion of planetesimals in
a protostellar nebula. In contrast to genuine {\it brown dwarfs}, defined in this review as gaseous objects of similar composition
as the parent cloud from which they formed by collapse. This issue is discussed in \S \ref{planetBD} and
observational diagnostics to differentiate brown dwarfs from planets, based on their different formation mechanisms, are suggested. Section \ref{conclusion} is devoted to the conclusion.

\bigskip
\section{\textbf{GASEOUS PLANETS: BIRTH AND EVOLUTION}}
\label{planet}

\bigskip
\noindent
\subsection{\textbf{ Planet formation}}
\label{planet_formation}
\bigskip

The conventional planet formation model is the core accretion model as developed by {\em Pollack et al.} (1996, hereafter P96). One of the major difficulties faced by this model is the long timescale necessary to form a gaseous planet like Jupiter, a timescale significantly larger than typical disk lifetimes, $\simle 10$ Myr. Reasonable timescales can be achieved only at the expense of arbitrary assumptions, like
e.g. nebula mean opacities reduced to 2\% of the ISM value in some temperature range or solid surface density significantly larger than the
minimum mass solar nebula value ({\em Hubickyj et al.}, 2005). 
This leaves the standard core accretion model in an uncomfortable situation. This model has been extended
recently by {\em Alibert et al.} (2004, 2005, hereafter A05) by
including the effects of migration and disk evolution during the planet formation process.
The occurence of migration during planet formation is supported by the discovery of numerous extrasolar giant planets at very short distance to their parent stars, well within the so-called ice line, about 5 AU for the solar nebula conditions.
Below this limit, above ice melting temperature, the insufficient surface density of solids that will
form eventually the planet core, and the lack of a large reservoir of gas prevent in-situ formation of large gaseous planets.

Moreover,
inward migration of the planet should arise from angular momentum transfer due to gravitational interactions between the gaseous disc and the growing planet ({\em  Lin and Papaloizou}, 1986; {\em Ward}, 1997; {\em Tanaka et al.}, 2002).
Taking into account the migration of a growing planet solves the long lasting timescale problem of the core
accretion scenario. Indeed, when migration is included, the planet feeding zone never
becomes depleted in planetesimals. As a result, the so-called phase 1 (see P96), dominated by accretion of solid material, is lengthened whereas phase 2, dominated by gas accretion, is shortened appreciably. During the last so-called phase 3, runaway gas accretion occurs and the predominantly H/He envelope is attracted onto the core.  Phase 3 is very short compared to phases 1 and 2,
and phase 2 essentially determines the formation timescale of the planet. The planet can thus form now
on a timescale consistent with disk lifetimes, i.e. a few Myr for a Jupiter (see A05).

In the models of {\em Bodenheimer et al.} (2000a) and {\em Hubyckij et al.} (2005), which are based on the P96 formalism, the calculations proceed in 3 steps: (i) the planet is bounded by its Roche lobe ($R_p=R_L$) (or more precisely by Min($R_L,R_{acc}$) where $R_{acc}=GM/c_s^2$ is the accretion radius and $c_s$ the local sound velocity in the disk) so
that the temperature and pressure at the planet surface are the ones of the surrounding nebula.
Note that in P96 calculations, opacity of the nebula is a key ingredient; (ii)
the planet external radius is the one obtained when the maximum gas accretion rate is reached. In P96,
this value is fixed to $1\times 10^{-2}\,\mearth$ yr$^{-1}$. At this stage, the external conditions have
changed ($R_p<R_L$). Matter falls in free fall from the Roche lobe to the planet radius, producing a shock luminosity; (iii) once the planet reaches its
{\it predefined} final mass, the accretion rate is set to 0 and the boundary conditions become the
ones of a cooling isolated object, $L=4\pi \sigma R^2\te^4$ and $\kappa_RP_{ph}={2\over 3}g$, where $\kappa_R$ denotes the
mean Rosseland opacity. The planet surface radius is
essentially fixed by the accretion shock conditions (see e.g. Fig. 1d of {\em Hubickyj et al.}, 2005). This value, however, remains highly uncertain, as its correct determination would imply a proper treatment of the radiative shock.
In A05, phase (i) is similar to step (i) described above, except that the planet migration from an initial arbitrary location and the disk evolution are taken into account, so that the thermodynamic conditions of the surrounding nebula, as well as the distance to the star, and thus the planet Roche lobe radius, change with time.
The planet's final mass is set by the accretion rate limit,
and is thus not defined a priori. Note that, because of the disk evolution and/or the creation of a gap around the planet, the accretion rate limit is 1 to 2 orders of magnitude smaller than
the one in P96 at the end of phase (i) and reaches essentially 0 with time, a fact supported by 3D hydrodynamical simulations ({\em D'Angelo et al.}, 2003; {\em Kley and Dirksen}, 2005). Eventually the planet opens a gap when its Hill radius becomes equal to the disk density scale heigth and migration stops or declines until the disk is dissipated (see A05 for details). The planet radius cannot be defined precisely in this model as it results from the competing effects of gas
accretion and planet contraction with changing boundary conditions as the planet migrates inward and the disk evolves. In any event, the final stages of accretion are likely to occur within streams (see e.g. {\em Lubow et al.}, 1999), i.e. non-spherically and, as mentioned above, the planet final radius remains highly uncertain, at least in any 1D calculation.

The migration rate, in particular type I migration for low-mass planet seeds, remains an ill-defined parameter in these calculations.
The observed frequency of extrasolar planets implies a rate significantly smaller than estimates done for laminar disks ({\em Tanaka et al.}, 2002).
Numerical modelling of turbulent disks yields significantly reduced migration rates ({\em Nelson and Papaloizou}, 2004, see also {\em D'Angelo et al.}, 2003).
It has
been suggested recently that stochastic migration, i.e. protoplanets following a random walk through the disk due to
gravitational interaction with turbulent density fluctuations
in the disk, may provide a means of preventing at least some planetary cores from migrating into the central star due to type I migration ({\em Nelson}, 2005). Based on these arguments, and for lack of better determinations, A05 divide the aforementioned rate of Tanaka et al. by a factor 10 to 100. As noted by these authors, numerical tests show that, provided the rate is small enough to preserve planet survival, its exact value affects the extent of migration but {\it not} the formation timescale, nor the planet final structure and internal composition.

\bigskip
\noindent
\subsection{\textbf{Planet evolution}}
\bigskip

\subsubsection{\textbf{Non irradiated planets}}
\bigskip

\begin{figure*}
 \epsscale{2.2}
  \plottwo{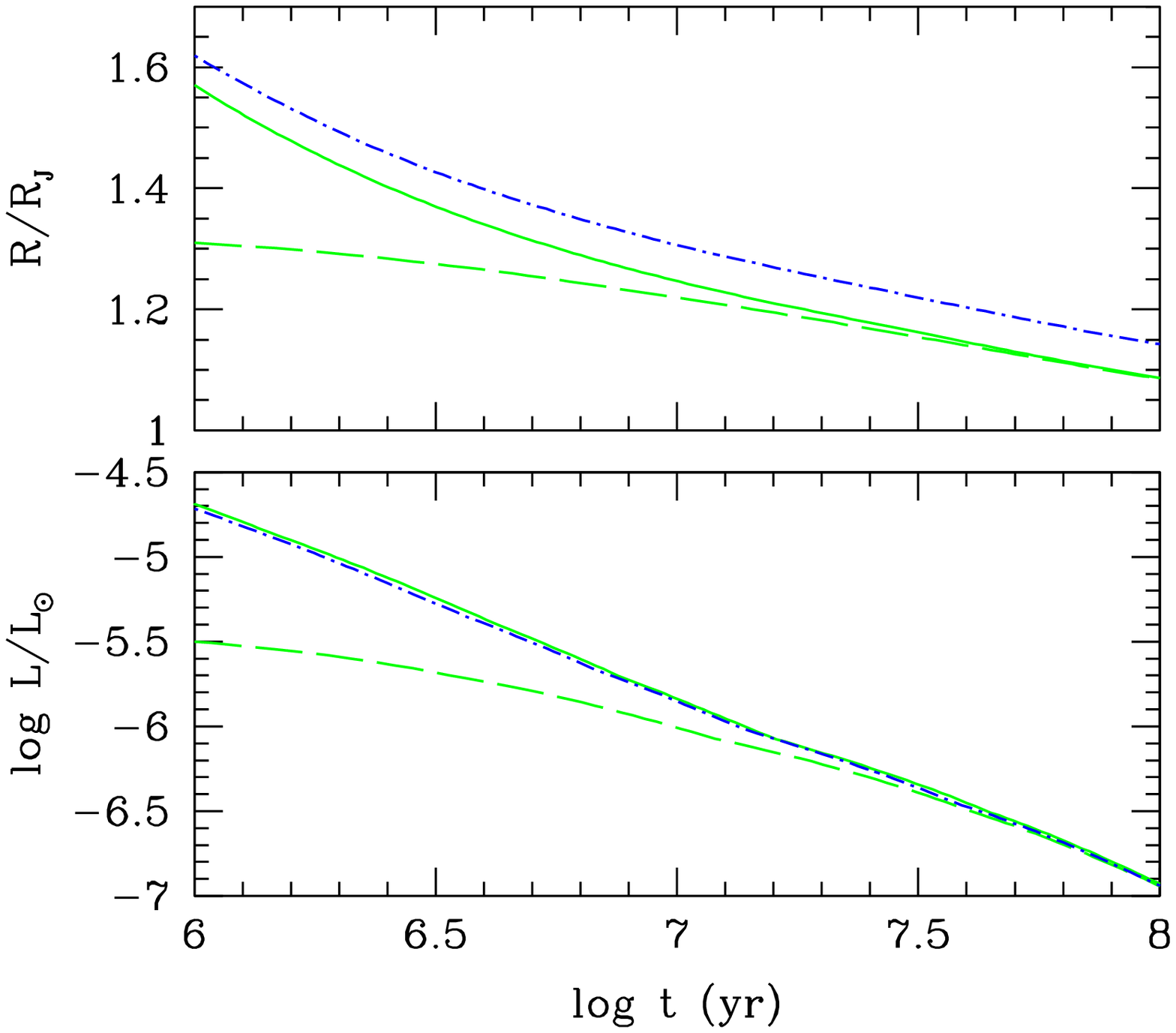}{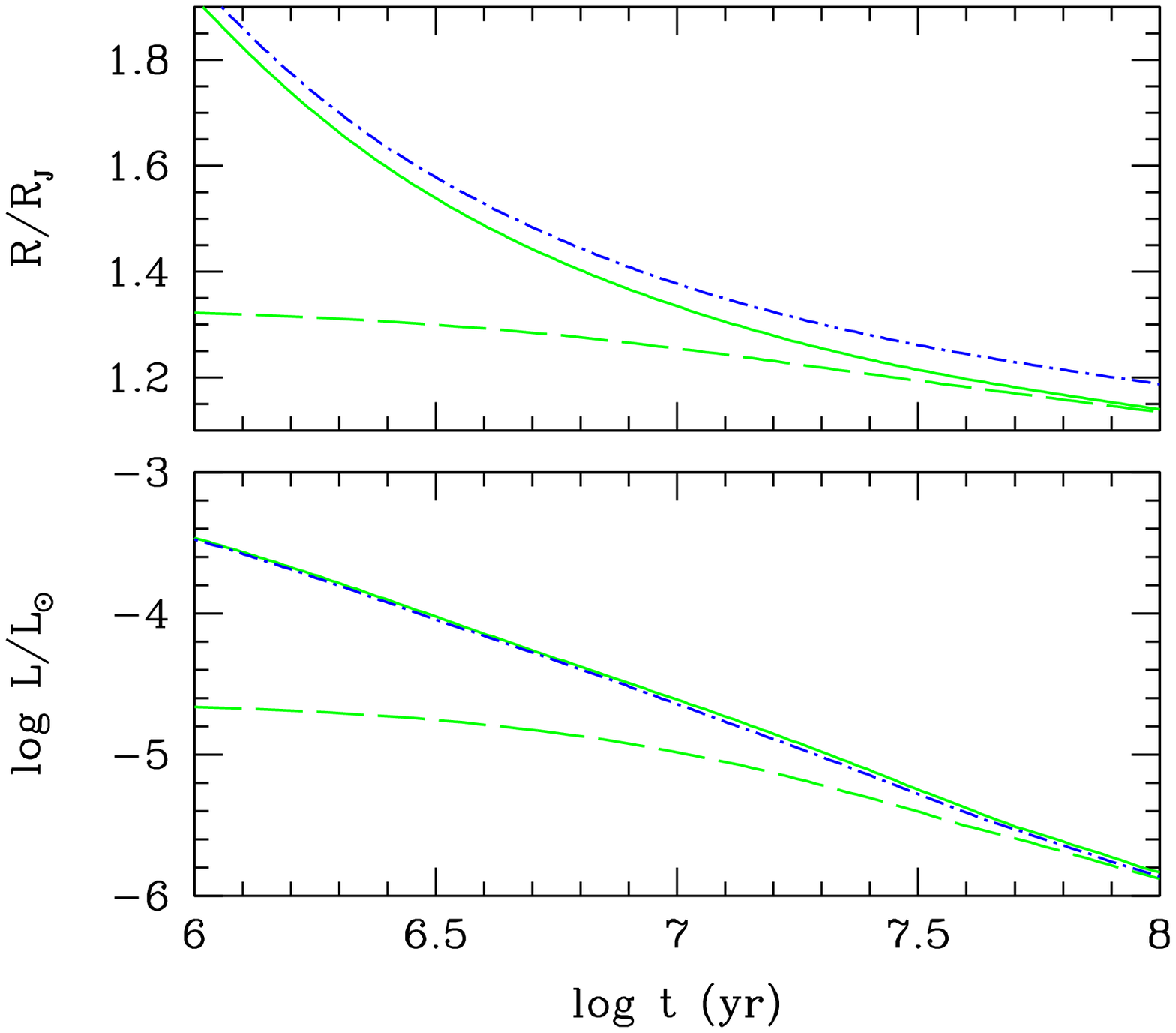}
  \caption{\small  Evolution of the radius and the luminosity for a 1 $\mjup$ (left) and a 4 $\mjup$ (right) planet with a 6 $\mearth$ solid core and $\mz/\menv$=10\%, for two different initial radii (solid vs dash lines, see text). The dot-dash lines portray the cooling of coreless, pure gaseous brown dwarfs of solar composition with similar initial radii as for the solid lines; the differences reflect the influence of the presence of a central core on the evolution.}
 \label{planetevol}
\end{figure*}

We first examine the evolution of young planets far enough from their parent star for irradiation effects to be neglected.
In order for the evolution to be consistent with the formation model, the planet structure includes now a central core
surrounded by an envelope enriched in heavy elements. These conditions are given by the formation model described in \S2.1, performed for different initial parameters (initial orbital distance, dust-to-gas ratio in the disc, photoevaporation rate, disc initial surface mass).
The planets are found to form with essentially the same core mass ($M_{\rm core} \simeq 6 \mearth$) independent of the planet final mass, whereas the heavy element mass fraction in the envelope deposited by the accreted planetesimals is found to increase substantially with decreasing total mass ({\em  Baraffe et al.}, 2006). The hydrogen-helium equation of state (EOS) is the Saumon, Chabrier and VanHorn EOS
({\em Saumon et al.}, 1995) whereas the thermodynamic properties of the heavy material relevant to the planet structure (ice, dunite($\equiv Mg_2SiO_4$), iron) are calculated with
the ANEOS EOS ({\em  Thompson and Lauson}, 1972). In the present calculations, we assume that the core is made of dunite, as representative of rock, yielding typical mean densities in the core
$\sim$ 6-7 g cm$^{-3}$. Comparative calculations with water ice cores, 
corresponding to a lower mean density $\sim$ 3 g cm$^{-3}$, change only slightly the mass-radius relationship for planets of identical core and total mass.
As mentioned above, the specific heat of the core is calculated with the ANEOS EOS so that the core contributes to the planet thermal evolution.
Fig. \ref{planetevol} displays the evolution of the radius and luminosity for 1 and 4 jupiter-mass planets, respectively. The solid and long-dash lines correspond to different initial radii for the new born planet, namely 3 and 1.3 $\rjup$ for the 1 $\mjup$ planet and 4 and 1.3 $\rjup$ for the 4 $\mjup$ planet, respectively. The 1.3 $\rjup$ case is similar to the calculations
of {\em Fortney et al.} (2005), based on the aforementioned formation model of {\em Hubickyj et al.} (2005). Note that these values are comfortably smaller than the Roche lobe limits at 5.2 AU from a Sun ($\simeq 530\,\rjup$
and $\simeq 830\,\rjup$ for a 1 $\mjup$ and a 4 $\mjup$ planet, respectively ({\em Eggleton}, 1983)).
The $t=0$ age for the planet evolution corresponds to the end of its
formation process, just after the runaway gas accretion (phase 3) has terminated. This planet formation timescale, namely $\sim$2-3 Myr, should thus be added to the ages displayed in 
Fig. \ref{planetevol} for the planet evolution. As seen in the figure, the difference between these initial conditions, namely a factor $\sim$2-3 in radius, affects the evolution of the planet for $10^7$ to $10^8$ yr, depending on its mass.
This reflects the significantly different thermal timescales at the begining of the evolution ($t=0$) for the different initial radii, namely $t_{KH}=GM^2/RL=3\times 10^5$ and $\sim5\times 10^7$ yr, respectively, for 1 $\mjup$. The smaller the initial radius the larger the consequences. Unfortunately, as mentioned above, uncertainties
in the models of planet formation prevent an accurate determination of the initial radius of the new born planet. Changing the maximum accretion rate or the opacity in P96, for example, or resolving the radiation transfer in the accretion shock, will very likely affect
the planet radius within a large factor.
Therefore, at least within the present uncertainties of the planet formation models, young gaseous
planets with cores and heavy elements in their envelopes can easily be 10 times brighter than suggested by the calculations of  {\em Fortney et al.} (2005) and thus are not necessarily "faint" in the sense that they can be as bright as pure gaseous, solar composition H/He objects of the same mass, i.e. low-mass brown dwarfs. In the same vein, the initial gravity of the planet can not be determined precisely and can certainly vary within at least an order of magnitude between
$\log g\sim 2$ and $\log g\sim 3$ for a jupiter-mass. Detections of young exoplanet luminosities with reasonable age 
determinations, i.e. within $\la 10$ Myr uncertainty, for instance in young clusters, would provide crucial information to help narrowing these uncertainties.


\bigskip
\subsubsection{\textbf{Effect of irradiation}}
\label{irrad}
\bigskip

We now examine the effects of irradiation on the evolution of close-in exoplanets, the so-called "hot-jupiters" and
"hot-neptunes" objects. Inclusion of the effect of irradiation of the parent star on the structure and evolution of
short period exoplanets has been considered by several authors. Only a few of these calculations, however,
are based on consistent boundary conditions between the internal structure and the {\it irradiated} atmosphere profiles. Such a proper boundary condition, implying consistent opacities in the atmosphere and interior structure calculations,
is determinant for correct evolutionary calculations of irradiated planets because of the growing external radiative zone which pushes the internal adiabat to deeper levels ({\em Guillot et al.}, 1996; {\em Seager and Sasselov}, 1998; {\em Barman et al.}, 2001, 2005). The out-going flux at the surface of the planet now includes the contribution from the incoming stellar flux ${\mathcal F_{\star}}$:

\begin{eqnarray}
{\mathcal F_{out}}&=& \sigma \teff^4+{\mathcal F_{inc}}=\sigma \teff^4+ f(\frac{R_{\star}}{a})^2 {\mathcal F_{\star}} \nonumber \\ &=& \sigma \teff^4+(1-A){\mathcal F_{inc}}+A{\mathcal F_{inc}}.
\label{Fout}
\end{eqnarray}

\noindent In Eq.\ref{Fout}, $\sigma \teff^4$ denotes the intrinsic internal flux of the planet, $A$ the Bond albedo and the last term on the r.h.s. of the equation is the reflected part of the spectrum. The factor $f$ is a geometrical factor characterizing the stellar flux redistribution over the planet surface
($f$= 1 implies the flux is redistributed over $\pi$ steradians, $f$=1/2 that it is redistributed over the day-side only, as intuitively expected for tidally locked planets, and $f$=1/4 over the entire planet surface). {\em Burkert et al.} (2005) have performed hydrodynamic calculations related to the heating of the night side of synchronously locked planets. With reasonable assumptions for the opacity in the atmosphere, these authors find that the
temperature difference between the day side and the night side could be in the $\sim 200$-300 K range, not enough to make an appreciable difference in the radius.
Previous estimates ({\em Showman and Guillot}, 2002; {\em Curtis and Showman}, 2005;
{\em Iro, B\'ezard and Guillot}, 2005), however, predict day/night temperature differences about
twice this value, and this issue needs to be further explored. From Eq. \ref{planetevol}, the evolution of the irradiated planet now reads:

\begin{equation}
L=-\int_MT\frac{dS}{dt}+4\pi  R_{p}^2\sigma T_{eq}^4+L_{reflected},
\label{Lout}
\end{equation}

\noindent where $T_{eq}^4={1-A\over \sigma}{\mathcal F_{inc}}=(1-A)f({R_\star \over a})^2T_\star^4$ denotes the planet equilibrium temperature, i.e. the temperature it would reach after exhaustion of all its internal heat content and contraction work ($\teff \rightarrow 0$).

As shown in {\em Chabrier et al.} (2004) and {\em Baraffe et al.} (2005), consistent calculations between the
irradiated atmospheric structure and the internal structure, which fixes the boundary condition for the planet photospheric radius, reproduce the radii of all observed transit planets so far,
without additional sources of internal heating,
except for HD209458b, which remains a puzzle (see Fig. 1 of {\em Baraffe et al.}, 2005). These calculations were based on planet interior models composed entirely of hydrogen and helium and do not include either a central core or heavy element enrichment in the envelope. The effect of a central rocky core on irradiated planet evolution has been examined by {\em Bodenheimer et al.} (2003)
but with simplified (Eddington) boundary conditions between the atmosphere and the interior. These authors found that for planets more massive than about 1 $\mjup$ the decrease in radius induced by the presence of a core is about 5\%, in agreement with previous estimates for non-irradiated planets ({\em Saumon et al.}, 1996). The effect, however, will be larger for less massive planets, including the recently discovered hot-Neptunes. This issue has been addressed recently by {\em Baraffe et al.} (2006), with proper, frequency-dependent atmosphere models. These authors find that, for a Saturn-mass planet ($\sim 100\,\mearth$), the difference in radius between a pure H/He planet and a planet with a 6 $\mearth$ core and a mass fraction of heavy element in the envelope $Z$=$\mz/\menv$=10\%,
as predicted by the formation model, is $R_Z/R_{HHe}\simeq 0.92$, i.e. a $\sim 9\%$ effect, possibly within present limits of detection.

A point of concern in the present calculations is that the boundary condition between the irradiated
atmospheric profile and the interior profile is based on atmosphere models of solar composition. Most of
the transiting planets, however, orbit stars that are enriched in metals and the planet atmosphere is supposed to
have the same enrichment. Calculations including such an enrichment are under work (see \S 5). The effect, however, is likely to be small for
two reasons. First of all, the enrichment of the parent stars remain modest, with a mean value $[M/H]\approx$0.2-0.3 ({\em Santos et al.}, 2004). Second of
all, irradiated atmospheric profiles display an extensive radiative zone (see above) so that gravitational settling may occur
even though, admitedly, various mixing mechanisms (e.g. decay of gravitational waves, convective overshooting, winds) could keep
gaseous heavy elements suspended in radiative regions.
Planets at large enough orbital distances for the effect of irradiation on the atmospheric
thermal profile to be negligible, however, should display significant
heavy element enrichment in their atmosphere, as observed for the giant planets of the solar system.

\bigskip
\subsubsection{\textbf{Evaporation}}
\label{evap}
\bigskip

The question of the long-term stability of gaseous close-in extrasolar giant planets has been raised since the discovery of 51 Peg b.
In the framework of Jeans approximation, the evaporation rate $\Phi$ (hydogen atoms cm$^{-2}$~s$^{-1}$) is given by
{\em Chamberlain and Hunten} (1987):

\begin{equation}
\Phi = \frac{n_{\mathrm{exo}}}{2\sqrt{\pi}}\sqrt{\frac{2\mathrm{k}T_{\mathrm{exo}}}{m}}\exp(-X) (1+X),
\end{equation}

\noindent where $n_{\mathrm{exo}}$ and $T_{\mathrm{exo}}$ are the number density and the temperature at the exobase (the level at which the mean free path of hydrogen atoms equals the scale height) and $X=v_{\infty}^{2}/v_{0}^2$ is the escape parameter, $v_{\infty}=(2GM_p/R_p)^{1/2}$ the planet escape velocity and $v_{0}=(2kT/m)^{1/2}$ the mean thermal velocity at $T_{\mathrm{exo}}$.
The first estimates of the evaporation rate of hot Jupiters (e.g. {\em Guillot et al.}, 1996) were obtained by using the equilibrium temperature $T_{eq}$  instead of the unknown value of $T_{\mathrm{exo}}$. For a typical 51Peg-b-like hot Jupiter, (1$\mjup$, $T_{\mathrm{eq}}\simeq$1300~K), the escape parameter  $X=v_{\infty}^{2}/v_{0}^2$ is then found to be larger than 150 whereas escape rates become significant for values below 20. On this basis, hot Jupiters were claimed to be stable over the lifetime of their star. However, $T_{eq}$ is not the relevant temperature for thermal escape, which occurs in the exosphere, where heating is due to XUV irradiation.
With simple assumptions, several authors estimated that the exospheric temperature could be of the order of 10,000~K ($X< 20$) and thus attempted the observation of the escaping H ({\em Moutou et al.}, 2001).  {\em Lammer et al.} (2003, L03) showed that the conditions allowing the use of Jeans approximation (hydrostatic equilibrium and negligible cooling by the escape itself) are not met in hot Jupiters, because of the considerable heating by stellar XUV. The application of Jeans escape yield unrealistically high exospheric temperatures ($X < 1$) in contradiction with the required hydrostatic hypothesis. They concluded that hot Jupiters should experience hydrodynamic escape, without a defined exobase,  where the upper atmosphere is continuously flowing to space and maintained at low temperature ($\ll 10,000~K$) by its expansion.
In this {\it blow-off} model, the escape rate of the main atmospheric component, H,  is only limited by the stellar XUV energy absorbed by the planet and is given by:

\begin{equation}
\dot M = 3 \left(\frac{\rm{R}_{XUV}}{\rm{R}_{p}}\right)^3 \epsilon {\mathcal F_\star} / ( G \rho),
\end{equation}

\noindent where $\rho$ is the mean planetary density and ${\mathcal F_\star}$ is the stellar flux, averaged over the whole planet surface, including both the contribution in the 1-1000~\AA{} wavelength interval and the 1215~\AA{} Lyman-$\alpha$ line. R$_{XUV}$ is the altitude of the (infinitely thin) layer where all the incoming XUV energy is absorbed while R$_p$ is the radius observed in the visible during a transit.  Here, $\epsilon$ would represent the heating efficiency, or the fraction of the incoming XUV flux that is effectively used for the escape. L03 applied a hydrodynamic model ({\em Watson et al.}, 1981) and estimated R$_{XUV}$/R$_{p} \approx 3$ for orbital distances closer than 0.1~AU. By assuming $\epsilon=1$ (or, in other words, that escape and expansion are the only cooling processes) they inferred the physical upper limit for the XUV-induced thermal escape rate to be $10^{12}$~g/s for HD209458~b at present time. Considering the evolution of XUV emission of main sequence G stars ({\em Ribas et al.}, 2005) and the significantly lower density of young gaseous planets implies rates 10 to 100 times higher in the early history of the hot Jupiters. Using these simple arguments, L03 suggested that hot Jupiters could have been initially much more massive although more detailed models are needed to better estimate the effective hydrodynamic escape rate.\\

Independently of this theoretical approach, {\em Vidal-Madjar et al.} (2003, VM03) measured the absorption in the Lyman-$\alpha$ line of HD~209458, using STIS onboard HST, during the transit of its planet. The decrease of luminosity they found is equivalent to the transit of a R$_{Ly\alpha}=3$~R$_p$ opaque disk. Although this observation seems to be consistent with L03, a larger but optically thin hydrogen cloud can also account for the observation. In fact, by noticing that the Roche lobe radius of the planet was 3-4~R$_p$, VM03 concludes that part of the observed hydrogen must consist in an escaping cometary-like tail. They estimated that the absorption implies an escape rate not lower than $10^{10}$~g/s.

The truncation of the expanded atmosphere by the Roche lobe, which was not considered by L03, has obviously to be taken into account in the mass loss process.  {\em Lecavelier et al.} (2004) proposed a {\it geometrical blow-off} model in which a hot exobase ($\sim10,000$~K), defined according to Jeans approximation, reaches the Roche lobe radius. This yields enhanced loss rates compared to a classical Jeans calculation that would not take into account the gravity field and the tidal distorsion of the atmosphere. {\em Jaritz et al.} (2005) argued that, although geometrical blow-off should occur for {\it some} of the known hot Jupiters, HD~209458b expands hydrodynamically up to 3~R$_p$ without reaching the L1 Lagrange point at which the Roche lobe overflow occurs. If confirmed, the debated observation of O and C in the expanded atmosphere of HD~209548~b ({\em Vidal-Madjar et al.}, 2004)  would favor the hydrodynamic regime, which is required to drag heavy species up to the escaping layers. However, the STIS instrument is no longer operational and similar observations will have to wait new EUV space observatories. Another indirect confirmation of the hydrodynamical regime is the absence of an H$_\alpha$ signature beyond R$_p$ ({\em Winn et al.}, 2004). This can be explained by the low temperature ($<5000$~K) expected in the hydrodynamically expanding atmosphere. {\em Yelle} (2004) published a detailed model of the photochemistry, radiative budget and physical structure of the expanding upper atmosphere of hot Jupiters and derived a
loss rate of $10^{8}$~g/s, about a factor 100 lower than the value inferred by VM03 from the observation. Recently, {\em Tian et al.} (2005) published an improved, multi-layer hydrodynamical model (compared to Watson), in which the energy deposition depth and the radiative cooling are taken into account. Rates of the order of $5\times10^{10}$~g/s are found, although they also depend on an arbitrary heating efficiency $\epsilon$. It is important to note that the composition of the expanding atmosphere in heavy elements can dramatically affect its behavior, mainly by modifying the radiative transfer (absorption and cooling).

Non-thermal escape is much more difficult to estimate as it depends on the unknown magnetic field of the planet  and stellar wind. Thermal escape is usually considered as the dominant mass loss process ({\em Grie{\ss}meier et al.}, 2004), but considering the complexity of the magnetic coupling between the star and the planet at orbital distances closer than 0.045~AU, unexpected non-thermal processes may still dominate the evaporation of some short-period exoplanets.

VM03 and L03 both suggested that the evaporation could lead to the loss of a significant fraction of the initial planetary mass and even to the evaporation of the whole planet, possibly leaving behind a dense core. In order to investigate the possible effects on the mass-radius evolution of close-in exoplanets,  {\em Baraffe et al.} (2004, 2005) included the maximum XUV-limited loss from L03 in the simulated evolution of a coreless gaseous giant planet, taking also into account the time dependency of the stellar XUV luminosity, calibrated on observations ({\em  Ribas et al.}, 2005). These studies showed that, even at the maximum loss rate, evaporation affects the long-term evolution of the radius only {\it below an initial critical mass}. For initial masses below this critical mass, the planet eventually vanishes in a very short but dramatic runaway expansion. This critical mass depends of course on the escape rate considered and drops to values much below 1 $\mjup$ when using lower rates like the ones predicted by Yelle, Tian et al., and Lecavelier et al.  ({\em Baraffe et al.}, 2006). One interesting result of the Baraffe et al. work
needing further attention is that evaporation does not seem to explain the surprisingly large visible radius (R$_p$) of HD~209458b, except if this planet is presently seen in its last and brief agony, which seems extremely unlikely.
The explanation for the large observed radius of HD~209458b thus remains an open question.

One may wonder whether this runaway evaporation phase can be studied with hydrostatic atmosphere models and quasi-static evolution models.
Atmospheric hydrostatic equilibrium is valid for values of the escape parameter $X>30$.
For a hot Jupiter at 0.045 AU, values of $X$ below 30 are found in the thermosphere, where the temperature
increases above 7000 K, at $R >$ 1.1 R$_p$ (see for instance {\em Yelle}, 2004). Such levels, with number densities $n<10^9$ cm$^{-3}$, lie
well above the levels where the boundary condition applies, i.e. near the photosphere with gas pressures $P\sim 10^{-5}$-10 bars.
The quasi-static evolution assumption is justified by the fact that, even though
the characteristic timescale of evaporation, $M/\dot M$, can become comparable to or even shorter than the Kelvin-Helmholtz timescale, $t_{\rm KH} \sim 2 G m^2/(RL)$,
it remains much larger than any hydrodynamical timescale. The present runaway phase, indeed, refers to a {\it thermal} runaway, like e.g. thermal pulses in AGB stars, characterized by a thermal timescale. Quasi-static evolution thus remains appropriate to study this
mass loss process, at least until truly hydrodynamic processes affect the planet photosphere.

More recently, Baraffe et al. (2006) examined the possibility for lower mass hot-neptune planets ($1\,\mnep=18\,\mearth\simeq0.06\,\mjup$) to be formed
originally as larger gaseous giants which experienced significant mass loss during their evolution. Depending on the value of the evaporation rates, these authors showed that presently observed (few gigayear old) neptune-mass irradiated planets may originate from objects of over a hundred earth masses if the evaporation rate reaches the maximum L03 value. For
$\sim 10$-20 times lower rates, as suggested e.g. by the hydrodynamical calculations of {\em Tian et al.} (2005), the hot-Neptunes would originate from objects of $\sim 50\,\mearth$, meaning that the planet
has lost more than 2/3 of its original mass. For rates a factor
100 smaller than L03, the effect of evaporation is found to become more modest  but a planet could still loose about 1/4 of its original mass due to stellar induced evaporation. These calculations, even though hampered by the large uncertainty in the evaporation rates, show that low-mass irradiated planets which lie below the afore mentioned critical initial mass, may have originally formed as objects with larger gaseous envelopes. This provides an alternative path to their formation besides other scenarios such as the core-collision model ({\em Brunini and Cionco}, 2005).

\bigskip
\section{\textbf{GRAVITATIONAL COLLAPSE OF PRESTELLAR CORES}}
\label{collapse}
\bigskip

After having examined the status of planet formation and evolution, we now turn to the formation and the early stages of evolution 
of stars and brown dwarfs.
In this section, we first review our current knowledge of the gravitational collapse of a protostar. We then will focus on the importance of non-spherical effects in the collapse.

\bigskip
\noindent
\subsection{\textbf{One dimensional models}}
\bigskip

Numerous authors have extensively considered the 1D collapse of a spherical cloud.
One of the most difficult aspects of the problem is the treatment of the cooling of the gas
due to collisional excitation of gas molecules, particularly during the late phase of the collapse
when the gas becomes optically thick. Radiative transfer  calculations coupled to hydrodynamics 
are then required. However, as noted originally by {\em Hayashi and Nakano} (1965) and confirmed by 
various calculations ({\em Larson}, 1969; {\em Masunaga and Inutsuka} 1998; {\em Lesaffre et al.}, 2005) 
the gas remains nearly isothermal for densities up to 10$^8$-10$^9$ cm$^{-3}$, making the 
isothermal assumption a fair and attractive simplification.

\bigskip
\noindent
\subsubsection{\textbf{The isothermal phase}}
\bigskip

The isothermal phase has been extensively investigated both numerically and analytically.
In particular, a family of self-similar solutions of the gravitational contraction has been 
studied in detail by {\em Penston} (1969), {\em Larson} (1969),  {\em Hunter} (1977), {\em Shu} (1977) 
and {\em Whitworth  and Summers} (1985). As shown by these authors, there is 
a 2D continuous set of solutions (taking into account the solutions which present
weak discontinuities at the sonic point) determined for example by the
value of the central density with bands of allowed and forbidden values. Two peculiar cases have 
been carefully studied, the so-called Larson-Penston and Shu solutions. The first case
presents supersonic velocities (up to 3.3 $c_s$ for large radius, where $c_s$ is the isothermal sound velocity) 
and is representative of very dynamical collapses. The second case assumes a quasistatic 
prestellar phase so that, at $t=0$, the density profile corresponds to the singular isothermal
sphere (SIS) and is given by $\rho _{SIS} \simeq c_s^2 / 2 \pi G r^2$.
 A rarefaction wave which propagates outwards is launched and the collapse is inside-out.
For both solutions the outer density profile is $\propto r^{-2}$ whereas in the neighbourhood 
of the central singularity, the density is  $\propto r^{-1.5}$.

Although the self-similar solutions depart significantly from the numerical calculations, 
they undoubtedly provide a physical hint on the collapse and the broad features described above 
appear to be  generic and are observed in the simulations. Following the work of 
{\em Foster and Chevalier} (1993), various studies have focussed on the collapse of a nearly critical 
Bonnor-Ebert sphere ({\em Ogino et al.}, 1999; {\em Hennebelle et al.}, 2003).  
This scenario presents a number of interesting features which agree well with observations 
of dense cores like those observed in the Taurus molecular cloud 
({\em Tafalla et al.}, 1998; {\em Bacmann et al.}, 2000; {\em Belloche et al.}, 2002). Namely:
 (i) the density profile is approximately flat in the centre during the prestellar phase; 
(ii) during the prestellar phase there
are (subsonic) inward velocities in the outer layers of the
core, whilst the inner parts are still approximately at rest; (iii) there
is an initial short phase of rapid accretion (notionally the Class 0 phase),
followed by a longer phase of slower accretion (the Class I
phase). This last feature is an important difference with the self-similar solutions, which have 
a constant accretion rate. The typical accretion rates obtained numerically are between the value 
of the Shu solution ($\dot{M}_{SIS} \simeq c_s^3/G$) and the Larson-Penston solution (about 50 $\times c_s^3/G$).

Motivated by the observations of much faster infall (see e.g. {\em Di Francesco et al.}, 2001),  triggered
collapses have  been considered ({\em Boss}, 1995; {\em Hennebelle et al.}, 2003, 2004; {\em Motoyama and Yoshida}, 2003). Much larger accretion rates, higher cloud densities and supersonic infall can be obtained in this context.
A close comparison between a strongly triggered collapse model and the class-0 protostar IRAS4A has been 
 performed with success by {\em Andr\'e et al.} (2004).

\bigskip
\noindent
\subsubsection{\textbf{Second Collapse and  formation of a young stellar object}}
\bigskip

When the  density becomes larger than $\simeq 10^{10}$ cm$^{-3}$ the gas becomes optically thick. The isothermal 
phase ends and the thermal structure of the collapsing cloud is nearly adiabatic. A thermally supported core forms ({\em Larson}, 1969; {\em Masunaga et al.}, 1998).
When matter piles up by accretion onto this hydrostatic core, its temperature and  density increase because of 
the stronger self-gravitating field. 
When the density of the first Larson core reaches about $10^{-7}$ g cm$^{-3}$, temperature is about 2000 K and the 
$H_2$ molecules start to dissociate ({\em Saumon et al.}, 1995). Most of the gravitational energy
goes into molecular dissociation energy so that
the effective adiabatic exponent, $\gamma=1+{d\, Ln T\over d\,Ln \rho}$ drops to about 1.1, significantly below the
critical value $\gamma$=4/3 ({\em Larson}, 1969; {\em Masunaga and Inutsuka}, 2000). Thermal pressure is therefore unable to support the hydrostatic core and the collapse restarts.

During the second collapse the temperature is roughly constant and close to
2000 K. When all the $H_2$ molecules have been dissociated into atomic 
hydrogen, the effective adiabatic exponent increases again above $\gamma$=4/3 and 
the star forms. The timescale of the second collapse is about the freefall
time of the first Larson core, $\sim$1 yr, very small compared with 
the timescale of the first collapse which is about 1 Myr. 

Both the first and second Larson cores are bounded during all the collapse 
of the cloud by an accretion shock in which the kinetic energy of the infalling material is
converted into heat.  
The effect of the accretion shock onto the protostar has been first 
considered by {\em Stahler et al.} (1980) and {\em Stahler} (1988).
The influence of accretion on the evolution of the protostar will be examined in \S \ref{accretion}.

\bigskip
\noindent
\subsection{\textbf{Influence of rotation and  magnetic field}}
\label{rot}
\bigskip

Here we examine the main influence of rotation and magnetic field 
on the cloud collapse, leaving aside 3D effects which are considered 
in \S 3.3.

\bigskip
\noindent
\subsubsection{\textbf{Effects of rotation}}
\bigskip

Rotation induces a strong anisotropy  in  the cloud, slowing  down 
and finally stopping  the equatorial material. {\em Ulrich} (1976) studied exact solutions for a rotating
and collapsing cold gas and showed that the equatorial density of the collapsing {\it envelope} is larger than in the 
absence of rotation. This has been further confirmed by {\em Terebey et al.} (1984) using an analytical 
solution which generalises the collapse of the SIS ({\em Shu}, 1977) in the case of a slowly rotating cloud. 
In the case of a $1\,\msol$ initially slowly rotating core  ($\beta=E_{rot}/E_{grav} \simeq 2 \%$), {\em Hennebelle et al.} (2004) estimate that the 
equatorial density of the collapsing envelope in the inner part of the cloud ($\simle 2000$ AU)
 can be 2 to 3 times higher than the axial one for a slow collapse and up to 10 times
higher in case of strongly compressed clouds.

The formation, growth and evolution of the rotationally supported disk has been modeled analytically by 
{\em Cassen \& Moosman} (1981) and {\em Stahler et al.} (1994). The growth of the disk drastically depends  on 
the angular momentum distribution, $j$. The centrifugal radius is about: $r_d \simeq j^2 / G M_{int}$
where $M_{int}$ is the mass inside the sphere of radius $r_d$.
Therefore,
for initial conditions corresponding to a  SIS in  solid body rotation,  $M_{int} \propto r$ and 
$j \propto r^2$, implying $r_d  \propto M_{int} ^3$.
On the contrary, starting with a uniform density sphere in solid body rotation, 
$M_{int} \propto r ^3$ and $r_d  \propto M_{int} ^{1/3}$,
 which implies much bigger disks. Such disks are indeed  found in hydrodynamical 
simulations of collapsing dense core initially in slow rotation.
For 1 $\msol$ dense cores with 
$\beta \simeq 2 \%$ the size of the disk during the class-0 phase is about 200 AU. 

The effect of the rotation on the forming protostar itself has been weakly explored. 
 2D equilibrium sequences of rotating protostars have been calculated by 
 {\em Durisen et al.} (1989).

\bigskip
\noindent
\subsubsection{\textbf{Effects of magnetic field}}
\bigskip

Magnetic field has been proposed to be the main support of the dense cores
against the gravitational collapse ({\em Shu et al.}, 1987) and the explanation for the 
low star formation efficiency in the Galaxy. Although this theory is now 
challenged by the origin of the support being mainly due to turbulence (see {\em Mac Low \& Klessen}, 2004 for a recent review), 
magnetic field certainly plays an important role in the formation of the protostar. 

The magnetically controlled collapse has been carefully investigated with 1D numerical 
simulations (e.g. {\em Mouschovias et al.}, 1985). 
It has been found that the collapse proceeds in 2 main phases, first a quasi-static contraction of the 
flattened cloud occurs through ambipolar diffusion and second, once a supercritical core has developed, it 
collapses dynamically. Quantitative estimates of the prestellar cloud lifetime are given in 
{\em Basu and Mouschovias} (1995). In strongly subcritical clouds (initial mass-to-flux ratio over critical 
mass-to-flux ratio smaller  than 1/10)  the formation of the protostar requires about 15 freefall times
whereas in a transcritical cloud (initial mass-to-flux ratio equal to critical mass-to-flux ratio), it 
requires about 3 freefall times. {\em Ciolek and Basu} (2000) showed that the collapse of the well studied 
prestellar cloud, L1544, is compatible with this core being transcritical. Note that, although the ambipolar diffusion time scale is much larger than the admitted star formation timescale, namely a few dynamical timescales, recent
2D simulations of compressible turbulence by {\em Li and Nakamura} (2004) suggest that enhanced ambipolar diffusion occurs through shock compression.

The transfer of angular momentum is  another important effect of magnetic fields.
It occurs through the emission of torsional Alfv\'en waves which carry away 
the angular momentum ({\em Mouschovias and Paleologou}, 1980; {\em Basu and Mouschovias}, 1995).
Since this process is more efficient if the rotation axis is perpendicular to the field lines (instead of parallel),
alignment between the magnetic field and the rotation axis is rapidly 
achieved. During the supercritical core formation epoch the angular velocity achieves a limiting profile
proportional to $1/r$ ({\em Basu}, 1998). Such a profile leads to centrifugal disks growing as
$r _d \propto M_{int}$ and thus intermediate between the very massive disks found in hydrodynamical 
simulations and the low-mass disks predicted by the SIS in solid body rotation model.

A very important difference between hydro and MHD cases is the presence of outflows in the 
latter ones, which have been found only recently in numerical simulations of collapsing protostellar core.
They are described in the next section.

Finally magnetic fields may induce a different mode of accretion.
Motivated by the observations of T Tauri stars, which are surrounded by a disk from which they
accrete material while having rotation velocities too small to be compatible with the conservation of angular momentum,
{\em K\"onigl} (1991) proposed
that most of the accreted matter may be channeled along the magnetic field lines from the disk 
to the poles of the star.  The angular momentum is then extracted from  the infalling gas
 by the magnetic field. The accretion onto the star occurring over a small fraction of its surface, 
significant differences with the case of spherical accretion are expected ({\em Hartmann et al.}, 1997),
an issue addressed in \S \ref{accretion}.

\bigskip
\noindent
\subsection{\textbf{Three dimensional models}}
\label{3D}

\begin{figure*}
 \epsscale{1.2}
  \plotone{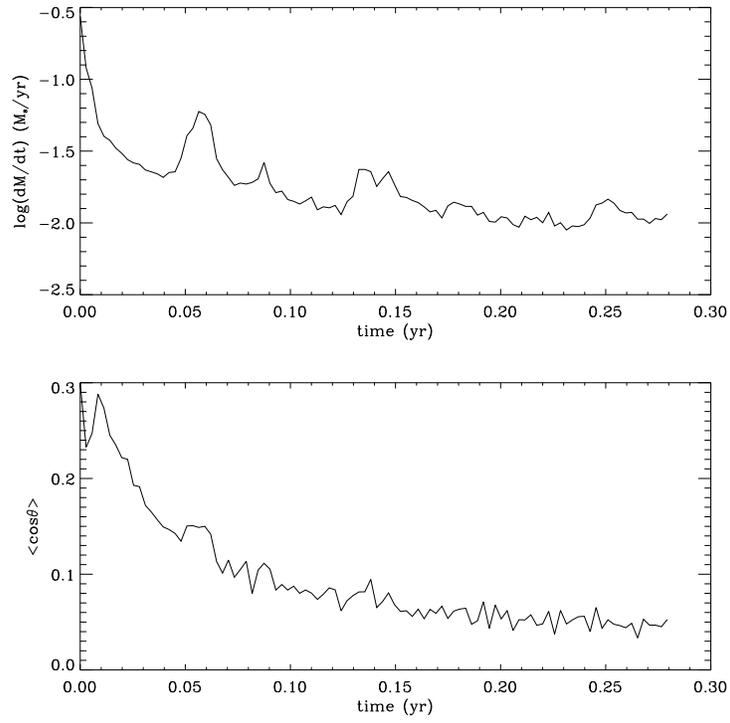}
\caption{\small Accretion rate (in $\msol$/yr) and average angle of accretion during the 3D simulation of the second collapse of a $10^{-3}\,\msol$ core.}
\label{3Daccret}
\end{figure*}

\begin{figure*}
 \epsscale{1.3}
   \plotone{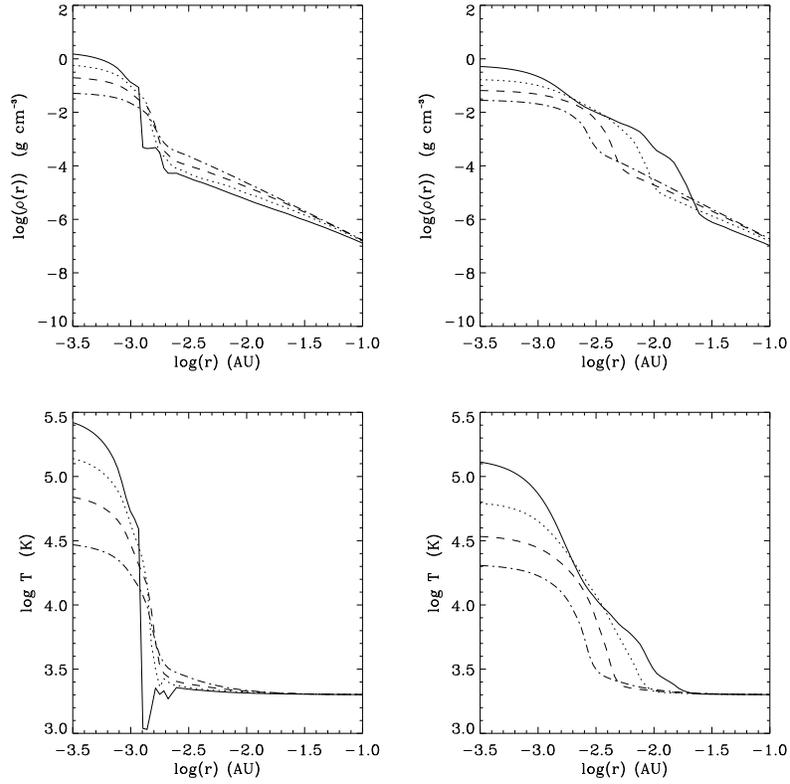}
\caption{\small Radial density and temperature profiles along the equatorial direction during the collapse of a $10^{-3}\,\msol$ core, for 4 different time steps.
 Left column: 1D (spherical) collapse (dash-dot=2.860 yr, dash=2.866 yr, dot=2,879 yr, solid=2.906 yr);
  right column: 3D collapse of a rotating core (dash-dot=3.594 yr, dash=3.606 yr, dot=3.634 yr, solid=3.704 yr).
  Note the different behavior of the accretion shock in the two cases.}
\label{3Dprofile}
\end{figure*}

\begin{figure}
 \epsscale{1.0}
  \plotone{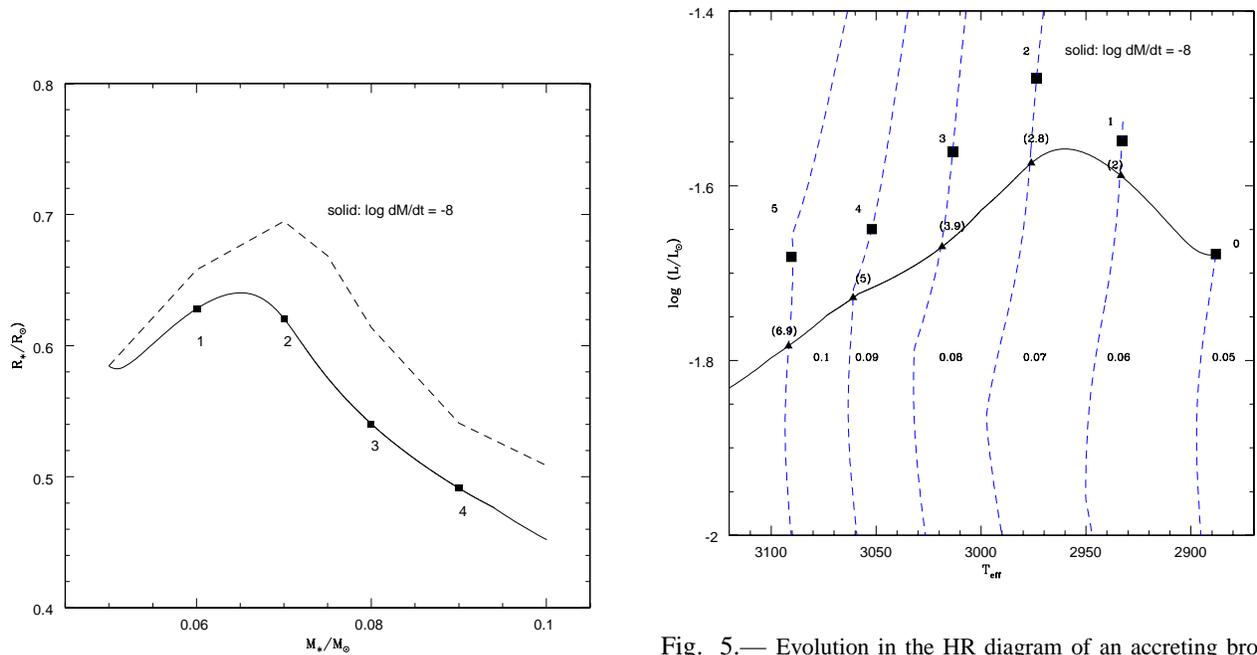}
  \caption{\small  Effect of accretion on the mass-radius relationship of
an accreting brown dwarf with initial mass 0.05 $\msol$ and accretion rate
$\dot M = 10^{-8} \msolyr$ (solid line). The dashed line indicates the radius of a non-accreting object with same mass and same age as its accreting counterpart.
Ages for the accreting object, in Myr, are indicated by the numbers.}
 \label{MR}
 \end{figure}

\begin{figure}
  \plotone{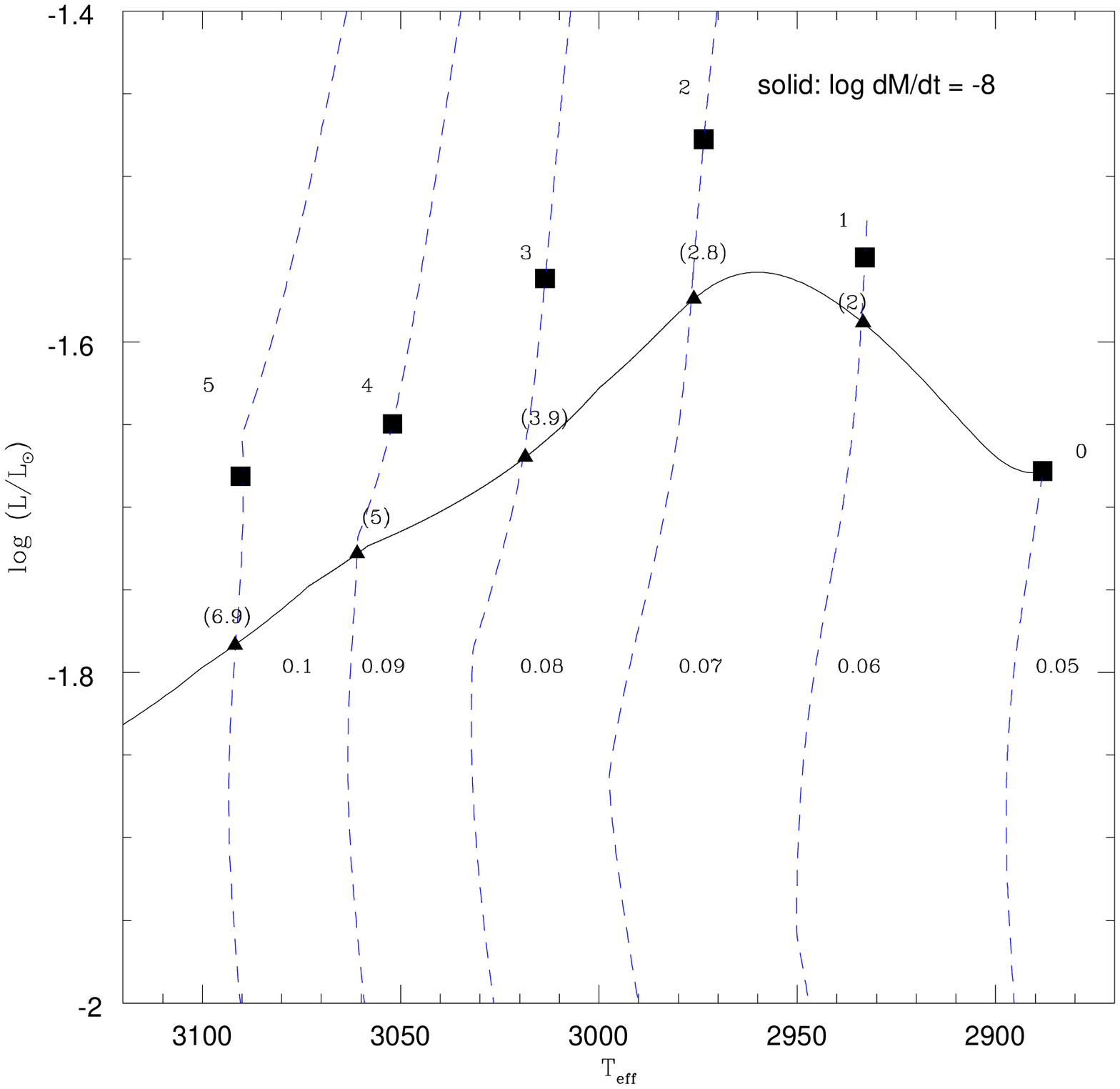}
  \caption{\small   Evolution in the HR diagram of an accreting brown dwarf, with initial mass 0.05 $\msol$ and accretion rate $\dot M = 10^{-8} \msolyr$ (solid line). The vertical dashed lines are cooling tracks of non-accreting low mass objects, with masses  indicated near the curves (from 0.05 $\msol$ to 0.1 $\msol$). The square symbols indicate
the position of non-accreting objects with the same age (indicated by the numbers
near the squares, in Myr) and same mass as the accreting counterpart (indicated by a triangle just below the corresponding square). The numbers in brackets (close
to the triangles) give the age (in Myr) of a non-accreting object  at
the position indicated by the triangle.}
\label{HRD}
 \end{figure}

\bigskip
\noindent
\subsubsection{\textbf{Axisymmetry breaking, transport of angular momentum
and fragmentation}}
\bigskip

One of the main new effects which appear in 3D calculations of a collapsing cloud is the 
axisymmetry breaking of the centrifugal disk. This occurs when its rotational energy
reaches about 40\% of its gravitational energy. Strong spiral modes develop
which exert a gravitational torque leading to a very efficient outwards  transport of angular momentum,
allowing  accretion onto the central object to continue. 
This effect has been modeled analytically ({\em Laughlin \& Rozyczka}, 1996)
and found by many authors in the numerical simulations (e.g. {\em Matsumoto and Hanawa}, 2003).

The fragmentation of the dense cores and the formation of multiple systems is one of the main
challenges of the field and entire chapters of  this book are dedicated to this subject. 
We refer to those as well as to the review of {\em Bodenheimer et al.} (2000b) for a 
comprehensive discussion of this topic.

\bigskip
\noindent
\subsubsection{\textbf{Multidimensional treatment of the second collapse}}
\bigskip

The second collapse leading to the formation of the protostar has been modeled in 
2 or 3D by various authors with two main motivations, namely modelling outflows and jets and 
explaining the formation of close binaries. Due to the large range of dynamical scales involved in the 
problem, the first calculations started from the first Larson core ({\em Boss}, 1989; {\em Bonnell and Bate}, 1994).
With the increase of computational power, calculations starting from prestellar core densities 
(e.g. 10$^{4}$ cm$^{-3}$) have been performed ({\em Bate}, 1998; {\em Tomisaka}, 1998; {\em Banerjee and Pudritz}, 2005).
For computational reasons,  the radiative transfer has  not been calculated self-consistently yet. Instead, 
piecewise polytropic equations  of state which mimics the thermodynamics of the cloud are often
 used ({\em Bate}, 1998; {\em Jappsen et al.},  2005).
More recently Banerjee and Pudritz (2005) used a tabulated cooling function which takes into account the 
microphysics of the gas with an approximated  opacity.

{\em Bonnell and Bate} (1994) conclude that fragmentation is possible during the second collapse. However 
since the mass of the stars is of the order of the Jeans mass, it is very small (0.01 $M_\odot$) and
 therefore they have to accrete most of their final mass. {\em Banerjee and Pudritz} (2005) form a close 
binary (with a separation of about 3 $R_\odot$) as well in their MHD adaptive mesh refinement
 calculations. Like Bate (1998) they find that
inside the large outer disk (60-200 AU) an inner disk of about 1 AU forms.

{\em Tomisaka} (1998) and {\em Banerjee and Pudritz} (2005)  report  outflows and jets during the collapse which 
contribute to carry away large amount of angular momentum. The physical mechanisms which 
is responsible for the launching of these outflows can be understood in terms of 
magnetic tower ({\em Lynden-Bell}, 2003). An annulus of highly wound magnetic field lines is 
created by the rotational motions and pushes the surrounding infalling material outwards.  
The physics involved in the jet is somehow different and based on the magneto centrifugal 
mechanism proposed by {\em Blandford and Payne} (1982).

In the 3D simulations below, we investigate inner core formation resulting from the collapse of a $10^{-3}\,\msol$ Bonnor-Ebert
sphere with densities and temperatures characteristic of the second core, namely $\rho \simeq 10^{-9}\,\gcc$ and $T\simeq 2000$ K.
We focus on the influence of tri-dimensional effects on the accretion geometry and on the inner profile of the core. Fig. \ref{3Daccret} displays the evolution of the accretion
rate $\dot M$ during the second collapse as well as the average angle of accretion
$ \langle \cos \theta \rangle$, i.e. the average angle between the vertical axis in spherical coordinates and the infalling gas. As seen in the figure, the accretion rate decreases immediately from a
large value close to the Larson-Penston prediction to a smaller Bondi-Hoyle or Shu like value, $c_s^3/G$, and accretion
occurs over a very limited fraction of the protostar surface, $ \langle \cos \theta \rangle < 0.3$ (spherical accretion would imply $ \langle \cos \theta \rangle=0.5$), so that most of the surface can radiate freely its
energy. This is important for the subsequent evolution of the object, as examined in the next section. The consequences of 3D effects on the density and temperature profiles of the protostar are illustrated in Fig. \ref{3Dprofile} which displays the equatorial density and
temperature profiles of the second Larson core at 4 time steps. Rotation leads to lower central densities and temperatures and to a more extended central core, as noted already by {\em Boss} (1989). These features are relevant for the internal energy transport - radiation vs convection - and the initial deuterium burning.
They also confirm that spherical collapse, although providing interesting qualitative information, cannot provide accurate initial
conditions for PMS tracks as it will overestimate (i) the internal temperature of the protostar and (ii) the surface
fraction covered by accretion,
thus preventing the object to contract at a proper rate.

\bigskip
\section{\textbf{EFFECT OF ACCRETION ON THE EARLY EVOLUTION OF LOW-MASS OBJECTS}}
\label{accretion}

\bigskip
\noindent
\subsection{\textbf{Observed accretion rates}}
\bigskip

Intensive investigations of accretion in young clusters and star formation regions show signatures of  this process over a wide range of masses, down to the substellar regime (see recent work by {\em Kenyon et al.}, 2005;  {\em Mohanty et al.}, 2005; {\em Muzerolle et al.},  2005, and references therein). In the youngest observed star forming regions, such as $\rho$-Ophiuchus with an age $\simle$ 1 Myr,  the fraction of accretors is greater than 50\%, independent of the mass ({\em Mohanty et al.}, 2005). This fraction decreases significantly with age, a fact interpreted as a decrease of the accretion rates below the observational limits, $\simle 10^{-12} \msol$/yr. The timescale for accretion rates to drop below such a measurable limit
is $\sim$ 5 Myr. In some cases, however, accretion continues up to $\sim$ 10 Myr. 
Note, however, that these age estimates for young clusters remain very uncertain, since they are usually based on evolutionary tracks that are not reliable at such ages ({\em Baraffe et al.}, 2002). Indeed, as demonstrated in {\em Baraffe et al.}, (2002), unknown initial conditions and unknown convection efficiency (mimicked in stellar evolution calculations by the mixing length parameters) during the early PMS contracting phase, characterized
by short Kelvin-Helmholtz timescales ($\simle 10^6$ yr), can affect drastically the contraction track of a young object in the Herzsprung-Rusell (HR) diagram. Therefore the age and/or mass of young objects can not be determined
accurately from observations, leading to very uncertain inferred disk lifetimes. However, even though the absolute timescales are uncertain, the trend of accretion rates decreasing with time is less questionable.
A sharp decrease of accretion rates with mass is also observed, with a correlation
$\dot  M \, \propto \, M^2$, all the way from solar mass stars to the smallest
observed accreting brown dwarfs, i.e $\sim$ 0.015 $\msol$
({\em Muzerolle et al.}, 2005). Typically, in the low mass star regime ($M \, \sim 0.2-1 \, \msol$), the accretion rates vary between $10^{-10} \msolyr$ and  $10^{-7} \msolyr$, whereas
below $\sim$ 0.2 $\msol$ and down to the BD regime, accretion rates range from $\sim  \, 5 \times 10^{-9} \, \msolyr$ to $10^{-12} \msolyr$ ({\em Muzerolle et al.}, 2003; {\em Natta et al.}, 2004;  {\em Mohanty et al.},  2005).
Last but not least, observations now show similarities of accretion properties between higher mass stars and low mass objects, including brown dwarfs, suggesting that stars and brown dwarfs share similar formation histories.

\bigskip
\noindent
\subsection{\textbf{Modeling the effect of accretion in young objects}}
\bigskip

On the theoretical front, {\em Stahler} (1983, 1988) has investigated the effect of {\it spherical accretion} onto protostars, defining the concept of a birth line,
a locus in the HR diagram where young stars first become optically visible when accretion ends.
Stahler  suggested that when the infall of material onto the protostar, responsible for its obscuration, ceases abruptly, the central object becomes an optically bright T Tauri star.

Since this benchmark work, progress in the observations of young objects have now shown that accretion occurs rapidly through a {\it disk}, as discussed in \S \ref{rot} and \ref{3D} and illustrated in Fig. \ref{3Daccret}.
The timescale for disk accretion is much longer than the strongly embedded protostellar phase, as
illustrated by the short lifetime of the class-0 objects compared with class-I.  Several studies have investigated the effect of accretion geometry on evolutionary tracks for low-mass and high-mass stars. These calculations
generally assume that (i) accretion takes place over a small fraction $\delta$ of the stellar surface and 
(ii) a dominant fraction of the accretion luminosity is radiated away and thus does not modify the
protostar internal energy content ({\em Mercer-Smith et al.},  1984; {\em Palla and Stahler}, 1992; {\em Hartman et al.}, 1997; {\em Siess et al.}, 1997), in contrast to the assumptions of {\em Stahler} (1988).
Under these conditions,
the luminosity of the accreting object is given by:

\begin{eqnarray}
L &=& \delta \cdot L_{acc}+L_D\nonumber \\ &-& (1-\delta)\int_M \Big\{T({dS\over dt})_m-T({\partial S\over \partial m})_{_t}{\dot m} \Big\}dm^\prime\,\,\,\,\,\,
\label{accret}
\end{eqnarray}

\noindent On the right hand side of Eq.\ref{accret}, the first term is the accreted luminosity, supposed to be entirely radiated away, $L_D$ is the D-burning luminosity, including freshly accreted deuterium, while the last term stems from the extra entropy at constant time due to the accreted mass (where ${\dot m}\equiv {\dot m}(m^\prime)$ is the accreted rate per mass shell).
The first assumption (i) is indeed relevant for thin disk accretion from a boundary layer or for magnetospheric accretion where the gas falls onto the star following magnetic accretion columns. It implies that most
of the stellar photosphere can radiate freely and is unaffected by a boundary layer
or accretion shocks. 
The second assumption (ii) depends on the details of the accretion process, which
remain very uncertain. In a attempt to study the impact of such an assumption on evolutionary models, one can assume that some fraction of the accreted matter internal energy is transferred to the protostar outer layers, the other fraction being radiated away. This extra supply of internal energy, per unit mass of accreted matter, is proportional to  the gravitational energy, $\epsilon GM/R$, with $\epsilon<1$ a free parameter.
As pointed out by {\em Hartmann et al.} (1997), the structure of an accreting object before or after ignition of deuterium, and the fact that it will be fully convective or will develop radiative layers, strongly depends on $\epsilon$, and to a lesser extent on assumption (i). For large values of $\epsilon$, convection can indeed be inhibited, even after deuterium ignition (see, e.g., {\em Mercer-Smith et al.}, 1984). 
Deuterium burning in the protostellar phase is also a central issue.
The key role played by deuterium burning on the properties of an accreting object 
and its location in the HR diagram was highlighted by {\em Stahler} (1983, 1988). Whether the deuterium fusion occurs in a fully convective object or in radiative layers is thus an important issue that affects significantly the structure of an accreting object.

Assuming that only a very small fraction of the thermal energy released by accretion
is added to the stellar interior, most of it being radiated away, {\em Hartmann et al.} (1997)  (see also {\em Siess et al.}, 1999) showed that, depending on its evolutionary stage, an accreting low mass star expands less or contracts more than a non accreting similar object. Consequently, an accreting object looks older in a HR diagram, because of its smaller radiating surface for the same internal flux, compared to a non accreting object at the same mass and age. This stems essentially from the accretion timescale becoming of the order of the Kelvin-Helmholtz timescale, for a given accretion rate, $M/\dot M\approx t_{KH}$, so that the contracting object does not have time to expand to the radius it would have in the absence of accretion.
An extension of these studies to the brown dwarf regime confirms these results,
in the case of significant accretion rate and no thermal energy addition due
to accretion ($\epsilon \, = \, 0$) ({\em Gallardo, Baraffe and Chabrier}, in preparation). Fig. \ref{MR} shows the effect of accretion on the radius of an object
with initial mass 0.05 $\msol$, with accretion rate $\dot M = 10^{-8} \msolyr$
and $\delta$ = 0, $\epsilon$ = 0. At any time, its structure is more compact than that of
a non accreting object of same mass (dashed curve in Fig. \ref{MR}), as mentioned above and as expected for accretion onto a fully convective object ({\em Prialnik and Livio}, 1985). The smaller radius, and thus the smaller luminosity, affect the location of the accreting brown dwarf in a HR diagram, as illustrated in Fig. \ref{HRD}. As seen on 
this figure,
assigning an age or a mass to an observed young object of a given luminosity using non-accreting tracks can significantly overestimate its age,
at least with the present accretion parameters. The effect of various accretion rates (see below) and of finite values
of $\epsilon$ is under study. This again illustrates the uncertainty
in age determination based on evolutionary tracks at young ages.

\bigskip
\noindent
\subsection{\textbf{Perspectives}}
\bigskip

The calculations presented above for an accreting brown dwarf have been done with no or small transfer of internal energy from the accretion shock to the brown dwarf interior. But our understanding of accretion mechanism is still too poor
to exclude  the release of a  large amount of energy due to accretion at deep levels. As mentioned previously, although current observations  indicate low accretion
rates ($\dot M \ll 10^{-8} \msolyr$) for brown dwarfs at ages $\gtrsim$1 Myr, they also point to rates decreasing with increasing time,
suggesting significantly larger accretion rates at early times ($\ll$ 1 Myr). If large amounts of matter are accreted, even through a disk, one expects a significant amount of thermal energy to be added to the object internal energy ({\em Hartman et al.}, 1997;  {\em Siess et al.}, 1999).
In which case we expect important modifications of the structure of the surface layers, with possible inhibition of convection as predicted for more massive objects ({\em Mercer-Smith et al.}, 1984; {\em Palla and Stahler}, 1992),
and thus a larger impact on ages and locations in the HR diagram than displayed in Fig. \ref{HRD}. Such effects of accretion need to be explored in details in order to get a better characterization of their impact on the early evolution of low mass stars and brown dwarfs
and thus of the uncertainties in mass and/or age determinations for young low mass objects. 
 
 \bigskip
\section{\textbf{BROWN DWARF VS PLANET: OBSERVABLE SIGNATURES}}
\label{planetBD}
 \bigskip


\begin{figure*}
 \epsscale{2.0}
  \plottwo{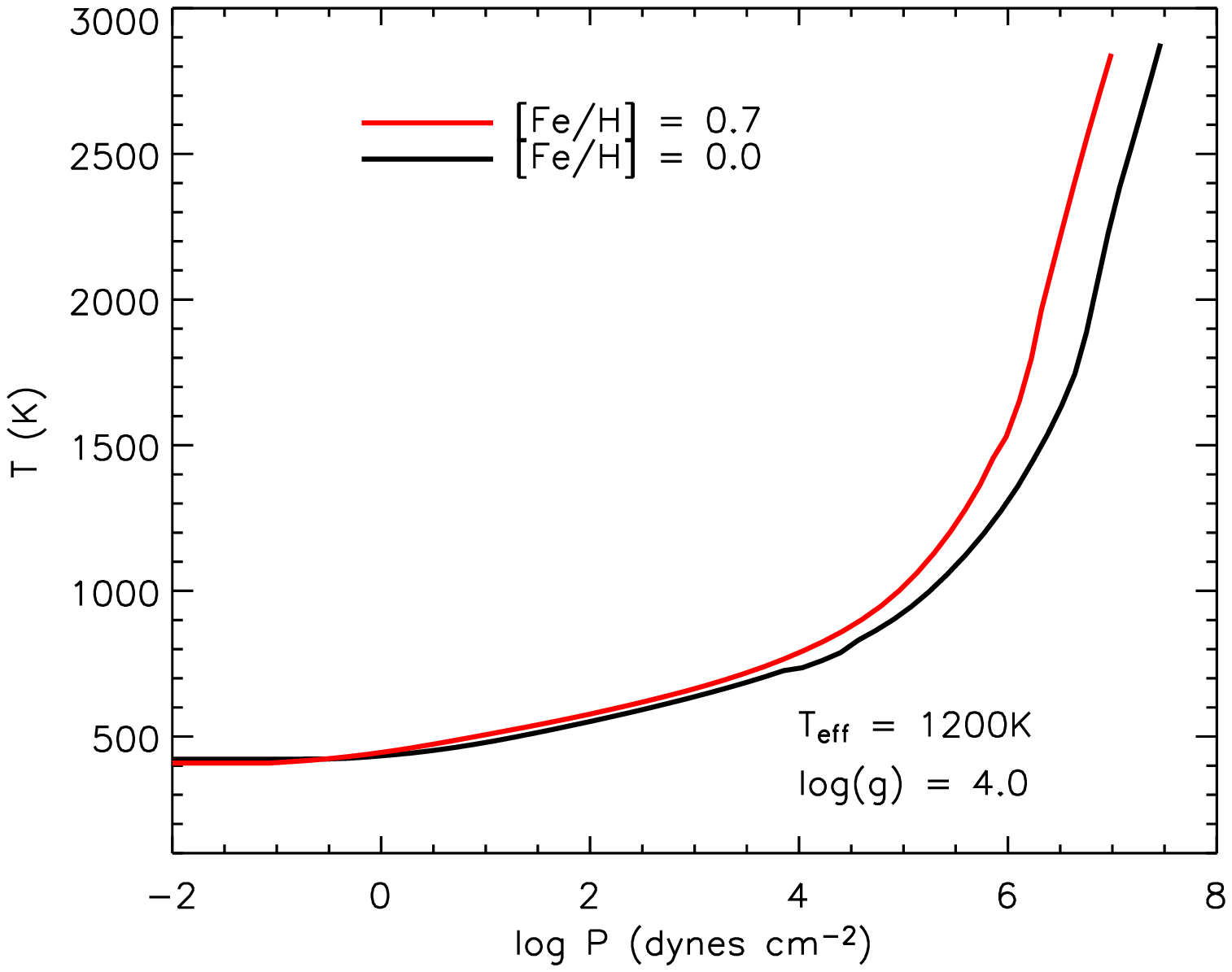}{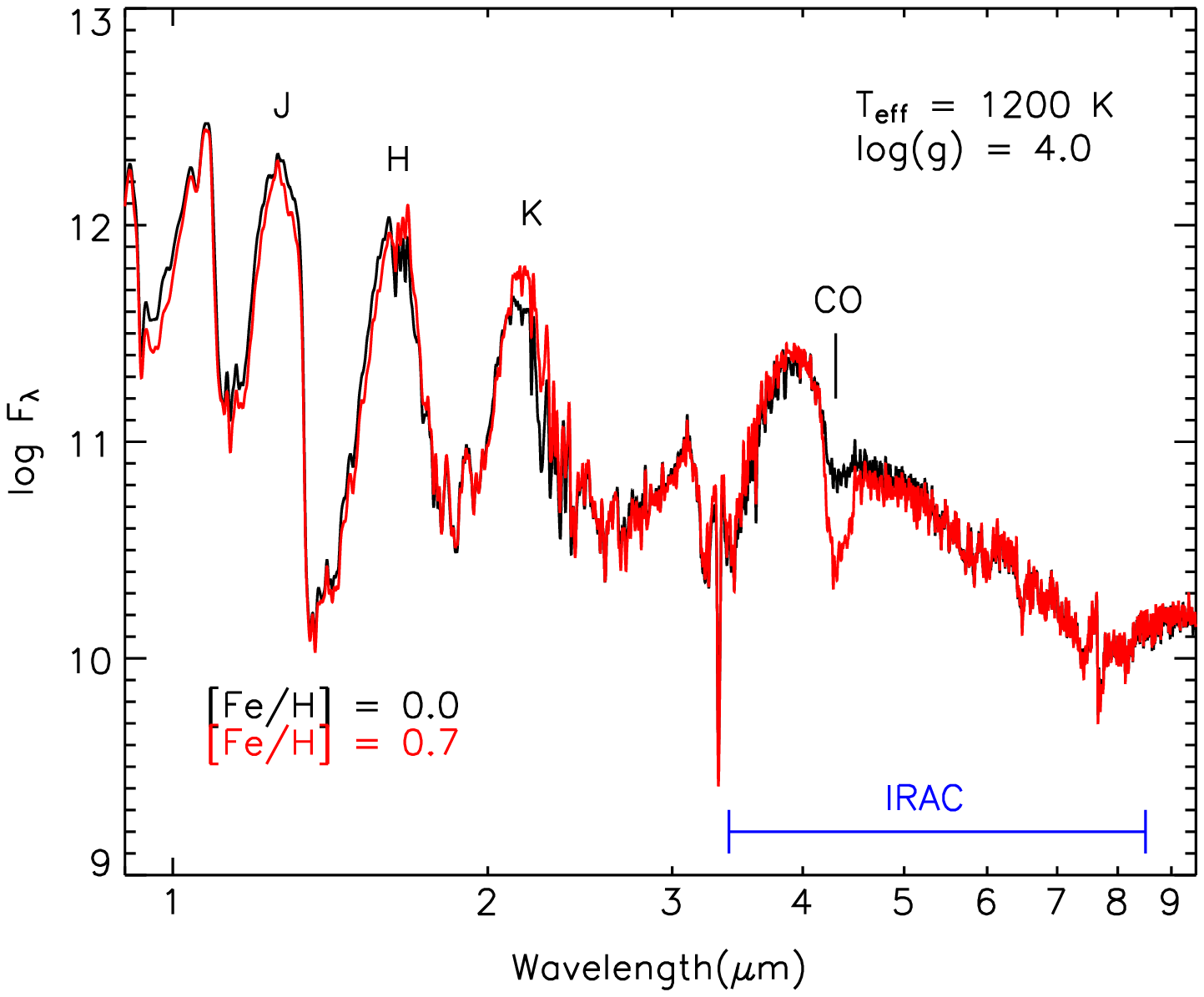}
  \caption{\small  Left: Temperature versus pressure for a young Jupiter-mass planet atmosphere model with
 solar and 5 times solar metal abundances (i.e., [Fe/H]=0.7). Right: Model spectra for the same conditions. The {\it Spitzer} IRAC filter is indicated.}
 \label{zstruc}
\end{figure*}

The "planetary status" of objects below the deuterium-burning limit, $\sim 13$ $\mjup$ ({\em Saumon et al.}, 1996; {\em Chabrier et al.}, 2000), remains
the subject of heated debate. The debate was recently intensified by the
direct image of an object below this mass limit, 2M1207b, orbiting a young brown dwarf at
a {\it projected} orbital distance $\gtrsim 55$ AU ({\em Chauvin et al.}, 2004). The present IAU working definition of a
``planet'' relies primarily on mass  -- not on the formation mechanism.
However, to understand the formation mechanisms of very low-mass objects, it is critical that we be
able to single out those which formed in a disk by a three step process as described in \S 2.1
(core-accretion followed by gas-capture) from low-mass, {\it no deuterium burning} objects which potentially formed by
gravitational collapse of a molecular cloud fragment.
According to the definition adopted in the present review, the former would be identified as genuine {\it planets}
while the latter would be {\it brown dwarfs}.
It is interesting, by the way, to note that D-burning is advocated to distinguish BDs from planets, whereas stars with masses below and above the limit for ignition of the CNO cycle share the same "star" denomination.
A common "brown dwarf" denomination should thus be used for D-burning or not D-burning BDs. Indeed,
D-burning is essentially inconsequential for the long term evolution of these objects, in contrast to steady hydrogen burning which yields nuclear equilibrium and determines completely
the fate of the object, star or brown dwarf (see e.g. {\em Chabrier and Baraffe}, 2000, Fig. 2 and 6).

In the coming decades, direct imaging surveys are certain to yield
a sizeable number of objects below $13$ $\mjup$ orbiting stars and brown dwarfs beyond a few
AU's -- a region unlikely to be well sampled by radial velocity surveys.  Without
a disk signpost, it will be difficult to distinguish long-period planets from very low mass brown dwarfs, based on
their different formation history. A very low-mass brown dwarf (that never burned deuterium) could well be mistaken
for a massive planet (see \S 2.2.1). Observable features that can
distinguish between these two types of objects are greatly needed.

Possible formation signatures could be contained in the atmospheric abundance
patterns of planets and their mass-luminosity relationships. As mentioned in \S 2.1,
a planet recently
forged in a disk by the three-step process will experience a brief period of
bombardment which enriches its atmosphere and interior in metals compared to its parent star abundances, as observed for our jovian planets ({\em Barshay and Lewis}, 1978; {\em Fegley and Lodders}, 1994; {\em B\'ezard et al.}, 2002, see also
the chapter by {\em Marley et al.}).
Brown dwarfs, on the other hand, should retain the abundance pattern
of the cloud from which they formed and,
in the case of BDs in binaries, should have abundances similar to their primary star.
The metallicity distribution of planet-hosting stars found by radial velocity
surveys already suggests that planet formation is favored in metal rich
environments thus making an {\em abundance test} even more attractive.  Recent
interpretations of {\em Spitzer} observations for two extrasolar planets, are
suggestive of non-solar C and O abundances (see the chapter by {\em Marley et al.} and references therein).


Enhanced metallicity leaves its mark on the interior, atmospheric structure, and
emergent spectrum in a variety of ways. As mentioned in \S \ref{irrad}, the presence
of a large heavy element content in the planet interior will affect its mechanical structure,
i.e. its mass-radius relationship. It will also modify its atmospheric structure.
Fig. \ref{zstruc} compares model atmospheric
structures for a young ($T_{eff} = 1200$K, $\log(g) = 4.0$), cloud-free, non-irradiated planet
mass object with solar and 5 times solar abundances.  As the atmospheric
opacities increase with increasing metallicity,
a natural warming occurs in the
deeper layers of the atmosphere.
This warming of the atmospheric structure will
have a direct impact on the evolution and predicted mass-luminosity
relationship.

\indent The right panel of Fig. \ref{zstruc} illustrates the spectral differences between these two models.
Clearly the most prominent effect is seen around 4.5 $\mu$m where the increased
absorption is due to an increase in CO.   Since this CO band falls in the {\em
Spitzer} IRAC (3 to 8 $\mu$m) coverage, significant metallicity enhancements in
planets could set them apart from typical brown dwarfs on an IRAC color-color
diagram.  There is also a noticeable increase in the $K$-band ($\sim 2.2 \mu$m)
flux.

The main purpose of this section is simply to point out one avenue to explore;
however, clearly a great deal of work must be done before a concise picture of
the expected abundance patterns in planets is developed. Non-equilibrium CO chemistry,
for example, is predicted to occur in cool so-called T-dwarf BDs ({\em Fegley and Lodders}, 1996; {\em Saumon et al.}, 2003). Moreover,
brown dwarfs forming by gravitational collapse will
certainly have abundance patterns as varied as their stellar associations, some
being relatively metal rich, e.g., the Hyades ({\em Taylor and Joner}, 2005). Additionally,
metallicity effects in broad band photometry could well be obscured by other
competing factors like gravity.  Also, our own solar system planets show a
range of C-O abundance ratios and varying levels of CO atmospheric enhancement
due to vertical mixing.  
Careful examinations of all these effects are necessary before any reliable spectral diagnostic can
be used to distinguish low-mass brown dwarfs from planets. Such a
diagnostic, however, has the virtue to rely on a physical distinction between two distinct populations in order to stop propagating confusion with improperly used "planet" denominations.

\bigskip
\section{\textbf{Conclusion}}
\label{conclusion}
\bigskip

In this review, we have explored (non exhaustively) our present understanding of the formation and the early
evolution of gaseous planets and protostars and brown dwarfs. We now have consistent calculations
between the planet formation, and thus its core mass and global heavy element enrichment, and the subsequent evolution after disk dissipation. These calculations are based on a revised version
of the core accretion model for planet formation, which includes planet migration and disk evolution,
providing an appealing scenario to solve the long standing timescale problem in the standard core accretion scenario. Uncertainties in the initial conditions of planet formation, unfortunately, lead to
large uncertainties in the initial radius of the new born planet. Given the dependence of the
thermal Kelvin-Helmholtz timescale on radius, this translates into large uncertainties on the characteristic luminosity
of young planets, over about $10^7$ yr for a $1\,\mjup$ planet.
Thus, it is impossible to say whether young planets are bright or faint and what is their initial gravity for
a given mass and therefore whether their evolution will differ from the one of young low-mass brown dwarfs. Conversely, future observations of young planets in disks of reasonably well
determined ages will enable us to constrain these initial conditions.

We have explored the effect of multidimensional collapse on the accretion properties and mechanical and
thermal structures of protostellar cores. These calculations demonstrate that, within less than a free fall time, accretion occurs
non-spherically, covering only a very limited fraction of the surface, so that most of the protostar surface
can radiate freely into space. Spherical collapse is shown to overestimate the inner density
and temperature of the prestellar core, yielding inaccurate initial conditions for PMS contracting tracks. This is important for initial deuterium burning and for energy transport, 3D inner structures having cooler temperatures
and more extended cores. This issue, however, can not be explored correctly with numerical tools available today
as it requires multidimensional implicit codes. The effect of accretion on the contraction of
young brown dwarfs was also explored.  Even though preliminary, these calculations confirm previous results for
pre-main sequence stars, namely that, for accretion timescales comparable to the Kelvin-Helmholtz timescale, the accreting object has a smaller radius than its non accreting counterpart, for the same
mass and age, and thus has a fainter luminosity. This smaller radius, along with the possible contribution from
the accretion disk luminosity, can lead to inaccurate determinations of young
object ages and masses from their location in an HR diagram, stressing further the questionable
validity of mass-age calibrations and disk lifetime estimates from effective temperature and luminosity
determinations in young clusters. These calculations also suggest that,
because of the highly non-spherical accretion, young stars or brown dwarfs will be visible shortly after
the second collapse and, depending on their various accretion histories, will
appear over an extended region of the HR diagram, even though being coeval. This seems to be
supported by the dispersion of low-mass objects observed in young stellar clusters or star forming regions when placed in an HR diagram (see e.g. Fig. 11 of {\em Chabier}, 2003). This
suggests that, in spite of all its merits, the concept of a well defined birth line is not a correct representation, as star formation rather leads to a scatter over an extended area
in the HR diagram.

Finally, we suggest the deuterium-burning official distinction between brown dwarfs and planets to be
abandoned as it relies on a stellar (in a generic sense, ie including brown dwarfs) quasistatic formation scenario which
now seems to be superseded by a dynamical gravoturbulent picture. Star formation and planet formation very likely overlap in the $\sim$ few $\mjup$ range and a physically motivated distinction between
these two different populations should reflect their different formation mechanisms. Within the general paradigm that brown dwarfs and stars form predominantly from the gravoturbulent
collapse of a molecular cloud and should retain the composition of the parent cloud and that planet form dominantly from planetesimal and gas accretion in a
disk and thus should be significantly enriched in heavy elements compared to their parent star, we propose that these distinctions
should be revealed by different mechanical (mass-radius) and spectroscopic signatures. Further
exploration of this diagnostic is necessary and will hopefully be tested by {\it direct} obervations of genuine exoplanets.
\bigskip

\textbf{ Acknowledgments.} The authors are very grateful to Willy Benz for useful discussions on planet formation and to the
anonymous referee for insightful comments.

\bigskip

\centerline\textbf{ REFERENCES}
\bigskip
\parskip=0pt
{\small
\baselineskip=11pt

\refs Alibert Y., Mordasini C. and Benz W. (2004)
{\em Astron. Astrophys., 417}, L25-28.

\refs Alibert Y., Mordasini C., Benz W. and Winisdoerffer C. (2005)
{\em Astron. Astrophys., 434}, 343-353 (A05).

\refs Andr\'e P., Bouwman J., Belloche A. and Hennebelle P. (2004)
{\em Astr. spa. sci., 292}, 325-337.

\refs Bacmann A., Andr\`e P., Puget J.-L., Abergel A., Bontemps S. and
and Ward-Thompson D. (2000)
{\em Astron. Astrophys., 361}, 555-580.

\refs Banerjee R. and Pudritz R. (2005)
astro-ph/0508374.

\refs Baraffe I., Chabrier G., Barman T.,  Allard F. and Hauschildt P. H. (2003)
{\em Astron. Astrophys., 402}, 701-712. 

\refs Baraffe I., Selsis F., Chabrier G., Barman T.,  Allard F. Hauschildt P. and Lammer H. (2004)
{\em Astron. Astrophys., 419}, L13-16.

\refs Baraffe I., Alibert Y., Chabrier G. and Benz, W. (2006)
{\em Astron. Astrophys.}, in press, astro-ph/0512091.

\refs Barman T., Hauschildt P. and Allard F. (2001)
{\em Astrophys. J., 556}, 885-895.

\refs Barman T., Hauschildt P and Allard F. (2005)
{\em Astrophys. J., 632}, 1132-1139.

\refs Barshay S.~S. and Lewis J.~S. (1978) {\em Icarus 33}, 593-611.

\refs Basu S. (1998).
{\em Astrophys. J., 509}, 229-237.

\refs Basu S. and Mouschovias T. C. (1995)
{\em Astrophys. J., 453}, 271-283.

\refs Bate M. (1998)
{\em Astrophys. J., 508}, L95-98.

\refs Belloche A., Andr\'e P., Despois D. and Blinder S., (2002)
{\em Astron. Astrophys., 393}, 927-947.

\refs  B{\'e}zard B., Lellouch E., Strobel D., Maillard J.-P. and Drossart P. (2002) {\em Icarus 159}, 95-111.

\refs Blandford R. D. and Payne, D. G. (1982)
{\em Mon. Not. Roy. Astr. Soc. 199}, 883-903.

\refs Bodenheimer P., Hubyckij O. and Lissauer J. (2000a)
{\em Icarus, 143}, 2-14.

\refs Bodenheimer P., Burkert A., Klein R. I. and Boss A. P. (2000b)
In {\em Protostars and Planets IV},
(Boss A. P., Russell S. S., eds.), Univ. of Arizon Press, Tucson, p.675.

\refs Bodenheimer P., Laughlin G. and Lin D. (2003)
{\em  Astrophys. J., 592}, 555-563.

\refs Bonnell I. A. and Bate M. R. (1994)
{\em Mon. Not. Roy. Astr. Soc. 271}, 999-1004.

\refs Boss A. P. (1989)
{\em  Astrophys. J., 346}, 336-349.

\refs Boss A. P. (1995)
{\em Astrophys. J., 439}, 224-236.

\refs Brunini A. and Cionco R.G. (2005)
{\em Icarus 177}, 264-268.

\refs Burkert A., Lin
D.~N.~C., Bodenheimer P.~H., Jones C.~A. and Yorke H.~W. (2005)
 {\em Astrophys. J., 618}, 512-523.

\refs Cassen P. and Moosman A. (1981)
{\em Icarus, 48}, 353-376.

\refs Chabrier G. (2003)
{\em Publ. Astron. Soc. Pac., 115}, 763-795.

\refs Chabrier G. and Baraffe I. (2000)
{\em Ann. Rev. Astron. Astrop. 38}, 337-377.

\refs Chabrier G., Baraffe I., Allard F. and Hauschildt, P.H. (2000) {\em Astrophys. J., 542}, L119-122.

\refs Chabrier G., Barman T., Baraffe I., Allard F. and Hauschildt P. H. (2004)
{\em Astrophys. J., 603}, L53-56.

\refs Chamberlain J. W. and Hunten D. M. (1987)
In {\em International Geophysics Series, Academic Press, Inc. 36}, 493.

\refs Chauvin G. et al. (2005)
{\em Astron. Astrophys., 438}, L25-28.

\refs Ciolek G. E. and Basu S. (2000)
 {\em Astrophys. J., 529}, 925-931.

\refs Curtis S. C. and Showman A. P. (2005)
{\em  Astrophys. J., 629}, L45-48.

\refs  D'Angelo G., Kley W. and Henning T. (2003)
 {\em Astrophys. J., 586}, 540-561.

\refs Di Francesco J., Myers P. C., Wilner D. J., Ohashi N. and Mardones D. (2001)
{\em  Astrophys. J., 562}, 770-789.

\refs Durisen R. H., Yang S., Cassen P. and Stahler S. W. (1989)
{\em Astrophys. J., 345}, 959-971.


\refs Eggleton P.P. (1983)
{\em  Astrophys. J., 268}, 368-369.

\refs Fegley B.~J. and Lodders K. (1994) {\em Icarus 110}, 117-154.

\refs Fegley B.~J. and Lodders K. (1996) {\em Astrophys. J., 472}, L37-40.

\refs Fortney J. J., Marley M. S., Hubickyj O., Bodenheimer P. and Lissauer J.J. (2005), astro-ph/0510009.

\refs Foster P. and Chevalier R. (1993)
{\em Astrophys. J., 416}, 303-311.

\refs Grie{\ss}meier J.M., Stadelmann A., Penz T. et~al. (2004)
{\em Astron. Astrophys., 425}, 753-762.

\refs Guillot T., Burrows A., Hubbard W.~B. Lunine, J.~I. and
	Saumon D. (1996)
{\em Astrophys. J., 459}, L35-38.

\refs Hartmann L., Cassen P. and Kenyon S. J. (1997)
{\em Astrophys. J., 475}, 770-785.

\refs Hayashi C. and Nakano T.  (1965)
{\em Progr. Theoret. Phys., 34}, 754.

\refs Hennebelle P., Whitworth A., Gladwin P. and Andr\'e P. (2003)
{\em  Mon. Not. Roy. Astr. Soc., 340}, 870-882.

\refs Hennebelle P., Whitworth A., Cha S.-H. and Goodwin S. (2004)
{\em  Mon. Not. Roy. Astr. Soc., 348}, 687-701.

\refs Hunter C. (1977)
{\em Astrophys. J., 218}, 834-845.

\refs Iro N., B\'ezard B. and Guillot T. (2005)
{\em Astron. Astrophys., 436}, 719-727.

\refs Jaritz G.F., Endler S., Langmayr D. et~al. (2005)
{\em Astron. Astrophys., 439},  771-775.

\refs Jappsen A.-K., Klessen R.~S., Larson R.~B., Li Y. and Mac Low M.-M. (2005)
{\em Astron. Astrophys., 435}, 611-623.

\refs Kenyon M.J., Jeffries R.D., Naylor T., Oliveira J.M. and Maxted P.F.L. (2005)
{\em Mon. Not. Roy. Astr. Soc., 356}, 89-106.

\refs Kley W. and Dirksen G. (2005)
{\em Astron. Astrophys.}, in press.

\refs K\"onigl A. (1991).
{\em Astrophys. J., 370}, 39-43.

\refs Lammer H., Selsis F., Ribas I. et~al. (2003)
{\em Astrophys. J., 598}, L121-124 (L03).

\refs Larson R. (1969)
{\em Astrophys. J., 145}, 271-295.

\refs Laughlin, G. and Rozyczka M. (1996)
{\em Astrophys. J., 456}, 279-291.

\refs Lecavelier des Etangs A., Vidal-Madjar A., McConnell, J.C. and H\'ebrard G. (2004)
{\em Astron. Astrophys., 418},  L1-4.

\refs Lesaffre P., Belloche A., Chi\`eze J.P. and Andr\'e P. (2005)
{\em Astron. Astrophys., 443}, 961-971.

\refs Li Z.-Y. and Nakamura F. (2004)
{\em Astrophys. J., 609}, L83-86.

\refs Lin D. and Papaloizou J. (1986)
{\em Astrophys. J., 309}, 846-857.

\refs Lubow S.~H., Seibert M. and Artymowicz P. (1999)
{\em  Astrophys. J., 526}, 1001-1012.

\refs Lynden-Bell D. (2003)
{\em  Mon. Not. Roy. Astr. Soc., 341}, 1360-1372.

\refs Mac Low M.-M. and Klessen R. S. (2004)
{\em Rev. Mod. Phys. 76}, 125-194.

\refs Masunaga H., Miyama S. and Inutsuka S.-I. (1998)
{\em  Astrophys. J., 495}, 346-369.

\refs Masunaga H. and Inutsuka S.-I. (2000)
{\em Astrophys. J., 531}, 350-365.

\refs Matsumoto T. and Hanawa T. (2003)
{\em Astrophys. J., 595}, 913-934.

\refs Mercer-Smith J.A., Cameron A.G. and Epstein R.I.   (1984)
{\em Astrophys. J., 279}, 363-366.

\refs Mohanty S., Jayawardhana R. and Basri G.  (2005)
{\em Astrophys. J., 626}, 498-522.

\refs Motoyama K. and Yoshida T. (2003)
{\em  Mon. Not. Roy. Astr. Soc., 344}, 461-467.

\refs Mouschovias T. Ch. and Paleologou E. V. (1980)
{\em  Astrophys. J., 237}, 877-899.

\refs Mouschovias T. Ch., Paleologou E. V. and Fiedler R. A. (1985)
{\em  Astrophys. J., 291}, 772-797.

\refs Moutou C., Coustenis A., Schneider J., et~al. (2001) {\em Astron. Astrophys., 371}, 260-266.

\refs Muzerolle J., Hillenbrand L., Calvet N., Briceno C. and Hartmann L. (2003)
{\em Astrophys. J., 592}, 266-281.

\refs Muzerolle J., Luhman K., Briceno C., Hartmann L. and Calvet N.   (2005)
{\em Astrophys. J., 625}, 906-912.

\refs Natta A., Testi L., Randich S. and Muzerolle J. (2004)
{\em Mem. S.A.It., 76}, 343-347.

\refs Nelson R. P. and Papaloizou J. 2004
{\em  Mon. Not. Roy. Astr. Soc., 350}, 849-864.

\refs Nelson R.P. (2005)
{\em Astron. Astrophys., 443}, 1067-1085.

\refs Ogino S., Tomisaka K. and Nakamura F. (1999)
{Pub. Astr. Soc. Jap., 51} 637-651.

\refs Palla F. and Stahler S. (1992)
{\em  Astrophys. J., 392}, 667-677.

\refs Penston M. (1969)
{\em Mon. Not. Roy. Astr. Soc. 145}, 457-485.

\refs Pollack J.~B., Hubickyj O., Bodenheimer P.,
	Lissauer J.~J., Podolak M. and Greenzweig Y. (1986)
        {\em Icarus, 124}, 62-85. (P96).

\refs Prialnik D. and Livio M. (1985)
{\em Mon. Not. Roy. Astr. Soc., 216}, 37-52.

\refs Ribas I., Guinan E.F., G{\"u}del M. and Audard M. (2005)
{\em Astrophys. J., 622}, 680-694.

\refs Santos N. C. et al. (2004) {\em Astron. Astrophys., 437}, 1127-1133.

\refs Saumon D., Chabrier G. and Van Horn H. M. (1995)
{\em Astrophys. J. Sup., 99}, 713-741.

\refs Saumon D., Hubbard W.~B., Burrows A., Guillot T.,
	Lunine J.~I. and Chabrier G. (1996)
	{\em Astrophys. J., 460}, 993-1018.

\refs Saumon D., Marley M.~S., Lodders K. and Freedman R.~S. (2003) {\em IAU Symposium 211} 345.

\refs Seager S. and Sasselov D.~D. (1998) 
{\em Astrophys. J., 502}, L157-160.

\refs Showman A. and Guillot T. (2002)
{\em Astron. Astrophys., 385}, 166-180.

\refs Shu F. H. (1977)
{\em Astrophys. J., 214}, 488-497.

\refs Shu F. H., Adams F. C. and Lizano S. (1987) 
{\em Ann. Rev. Astron. Astrophys., 25}, 23-72.

\refs Siess L., Forestini M. and Bertout C. (1997) 
 {\em Astron. Astrophys., 326}, 1001-1012. 

\refs Siess L., Forestini, M. and Bertout C. (1999). 
 {\em Astron. Astrophys., 342}, 480-491.

\refs Stahler S. W.  (1983)  
{\em Astrophys. J., 274}, 822-829.

\refs Stahler S. W.  (1988)  
{\em Astrophys. J., 332}, 804-825.

\refs Stahler S. W., Shu F. H. and Taam R. E. (1980)
{\em   Astrophys. J., 241}, 637-654.

\refs Stahler S. W., Korycansky D. G., Brothers M. J. and Touma, J. (1994)
{\em Astrophys. J., 431}, 341-358.

\refs  Tafalla M., Mardones D., Myers P. C., Caselli P., Bachiller R. and Benson P. J. (1998)
{\em Astrophys. J., 504}, 900-914.

\refs Tanaka H., Takeuchi T. and Ward, W.R. (2002)
{\em Astrophys. J., 565}, 1257-1274.

\refs Taylor B.~J. and Joner M.~D. (2005)
{\em Astrophys. J. Supp., 159}, 100-117.

\refs Terebey S., Shu F. H. and Cassen P. (1984)
{\em  Astrophys. J., 286}, 529-551.

\refs Thompson S.L. and Lauson H.S. (1972)
In {\em Technical Report SC-RR-61 0714, Sandia National Laboratories}.

\refs Tian F., Toon O.B., Pavlov A.A. and De Sterck H. (2005)
{\em Astrophys. J., 621}, 1049-1060.

\refs Tomisaka K. (1998)
{\em Mon. Not. Roy. Astr. Soc., 502}, L163-167.

\refs Ulrich R. K. (1976)
{\em Astrophys. J., 210}, 377-391.

\refs Vidal-Madjar A., Lecavelier des Etangs A., D{\'e}sert J.M. et~al. (2003)
  {\em Nature, 422}, 143-146.

\refs Vidal-Madjar A., D{\'e}sert J.M., Lecavelier des Etangs A. et~al (2004)
{\em Astrophys. J., 604}, L69-72.

\refs Watson A.J., Donahue T.M. and Walker J.C.G. (1981)
{\em  Icarus, 48}, 150-166.

\refs Winn J.N., Suto Y., Turner E.L.- et~al. (2004)
{\em PASJ, 56}, 655-662.

\refs Ward W.R. (1997)
{\em Astrophys. J., 482}, L211-214.

\refs Whitworth A. Summers D. (1985)
{\em Mon. Not. Roy. Astr. Soc., 214}, 1-25.

\refs Yelle R.V. (2004)
{\em Icarus, 170,} 167-179.

\end{document}